\documentclass[11pt]{article}
\pdfoutput=1
\usepackage{jheppub}
\usepackage{amsmath}
\usepackage{amssymb}
\usepackage{dcolumn}
\usepackage{bm}
\usepackage{blkarray}
\usepackage{slashed}                             
\usepackage{hyperref,graphicx}           
\usepackage{url}
\usepackage{caption}
\usepackage{subcaption}
\usepackage{multirow}
\usepackage{units}
\usepackage{physics}
\usepackage{xcolor}
\usepackage{graphicx}
\usepackage{tikz}
\usepackage{siunitx}[=v2]
\usepackage[all]{hypcap}
\usepackage{cleveref}
\usepackage[mathlines]{lineno}
\usepackage{tikz-3dplot}
\usepackage{color}
\usepackage{multirow}
\usepackage{float}
\usepackage{mathtools}

\graphicspath{{./Figure/}}
\graphicspath{{./../Figure/}}

\begin{document}
\title{\boldmath Jupiter missions as probes of dark matter}
\author[a]{Lingfeng Li,}
\author[a,b]{JiJi Fan}
\affiliation[a]{Department of Physics, Brown University, Providence, RI, 02912, USA}
\affiliation[b]{Brown Theoretical Physics Center, Brown University, Providence, RI, 02912, USA}
\emailAdd{lingfeng\_li@brown.edu}
\emailAdd{jiji\_fan@brown.edu}

\vspace*{1cm}

\abstract{Jupiter, the fascinating largest planet in the solar system, has been visited by nine spacecraft, which have collected a significant amount of data about Jovian properties. In this paper, we show that one type of the \textit{in situ} measurements on the relativistic electron fluxes could be used to probe dark matter (DM) and dark mediator between the dark sector and our visible world. Jupiter, with its immense weight and cool core, could be an ideal capturer for DM with masses around the GeV scale. The captured DM particles could annihilate into long-lived dark mediators such as dark photons, which subsequently decay into electrons and positrons outside Jupiter. The charged particles, trapped by the Jovian magnetic field, have been measured in Jupiter missions such as the Galileo probe and the Juno orbiter. We use the data available to set upper bounds on the cross section of DM scattering off nucleons, $\sigma_{\chi n}$, for dark mediators with lifetime of order ${\cal O}(0.1-1)$s. The results show that data from Jupiter missions already probe regions in the parameter space un- or under-explored by existing DM searches, \textit{e.g.}, constrain $\sigma_{\chi n}$ of order $(10^{-41} - 10^{-39})$ cm$^2$ for 1 GeV DM dominantly annihilating into $e^+e^-$ through dark mediators.  This study serves as an example and an initial step to explore the full physics potential of the large planetary datasets from Jupiter missions. We also outline several other potential directions related to secondary products of electrons, positron signals and solar axions. }
\maketitle
\flushbottom

\section{Introduction}
\label{sec:intro}

\begin{figure}[h!]
\centering
\includegraphics[width=10cm]{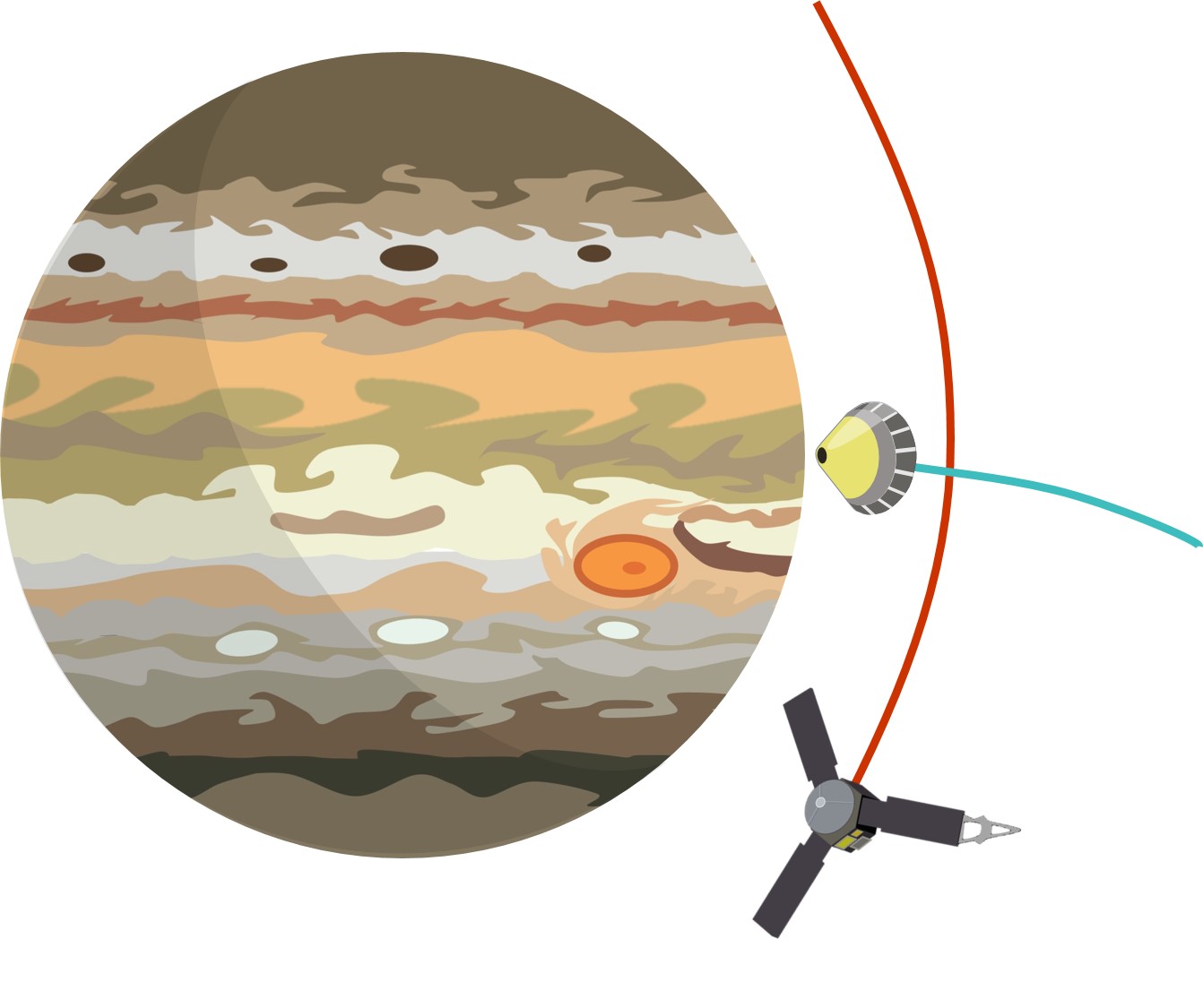}
\caption{A cartoon of the trajectories of the Galileo probe (blue) and the Juno orbiter (red). }
\label{fig:trajectory}
\end{figure}

Jupiter, fifth in line from the Sun, is the largest planet in our solar system with a weight more than twice that of all the other planets combined. Its fascinating properties, including the vivid stripes and swirls on the surface, the great red spot as a giant storm about twice the size of the earth, and more than 75 moons, make it a target of continuous investigation and exploration, which dates all the way back to at least the Babylonian astronomers in the 7th or 8th century BC. In modern times, Jupiter has been visited by nine spacecraft, collecting a plethora of information about the giant. Among them, seven just flew by, such as the Pioneer~\cite{1974Sci...183..301H, 1975Sci...188..447O} and Voyager missions~\cite{1979Sci...204..945S, 1979Sci...206..925S}, while two have orbited Jupiter, the Galileo mission~\cite{1992SSRv...60....3J} and the Juno mission~\cite{2017SSRv..213....5B}. The flybys provided snapshots of Jupiter. On the other hand, the Galileo mission entered the Jovian magnetosphere and released the Galileo probe, which dived into the atmosphere~\cite{young1998galileo, 2000JGR...10512093Y}, while the Galileo orbiter remained and orbited within the Jovian radiation belts and provided an extensive survey of the belts. The latest Juno mission is the second one in NASA's new frontiers program. The spacecraft, launched in 2011, was inserted into the orbit around Jupiter in 2016 and will continue its investigation till 2025 (or its end of life). The trajectories of the Jupiter missions, Galileo probe and Juno orbiter, are depicted in Fig.~\ref{fig:trajectory}.

No doubt that Jupiter is an important object in astronomy and planetary science. More intriguingly, data collected by the Jupiter missions could contribute to an apparently unrelated scientific endeavor: the hunt for dark matter (DM) and the dark sector, which will be our focus in this article. It has already been pointed out that Jupiter could be a powerful DM capturer~\cite{Kawasaki:1991eu, Adler:2008ky, Leane:2021ihh, Leane:2021tjj}: DM particles in the galactic halo could be captured by Jupiter if they scatter with the Jovian matter and lose enough kinetic energy so that they become gravitationally bound and accumulate inside Jupiter. Compared to other planets in the solar system, e.g., our Earth, the gas giant could capture more DM, enhancing the annihilation signals of captured particles. Compared to the Sun, Jupiter is much cooler in the core and the captured DM could remain inside, ideal for searches of (sub-)GeV DM which evaporate away even if captured initially by the Sun~\cite{Leane:2021tjj}. 

So far, the only detection proposal of DM captured in Jupiter is to search for gamma rays~\cite{Leane:2021tjj}. The class of dark sector models being probed is as follows: DM particles annihilate into a pair of dark mediators, which are portals between the dark sector and our visible sector, which subsequently decay into two gamma-ray photons outside Jupiter, which could be searched for using 12 years data of {\it Fermi} Large Area Telescope (LAT)~\cite{2009arXiv0907.0541G}, which is close to the Earth.\footnote{DM capture by celestial objects was proposed and computed first in \cite{Press:1985ug,Gould:1987ir,Gould:1987ww,1990ApJ...356..302G}. In the context of dark sector models with long-lived dark mediators, studies of signals from DM capture by other celestial objects such as the Sun and the Earth have been implemented in~\cite{Liu:2008kz, Batell:2009zp, Schuster:2009au, Schuster:2009fc, Bell:2011sn, FermiLAT:2011ozd, GarciaGarcia:2015fol, Feng:2015hja, Feng:2016ijc, Kouvaris:2016ltf, Allahverdi:2016fvl,Brdar:2016ifs, Smolinsky:2017fvb, Ardid:2017lry, Leane:2017vag, Arina:2017sng, Robertson:2017hdw, HAWC:2018szf, Nisa:2019mpb, Niblaeus:2019gjk, Cuoco:2019mlb, Mazziotta:2020foa, Bell:2021pyy, Bose:2021yhz, Leane:2021ihh, Zakeri:2021cur, Bose:2021cou}.}  

\begin{figure}[h!]
\centering
\includegraphics[width=10cm]{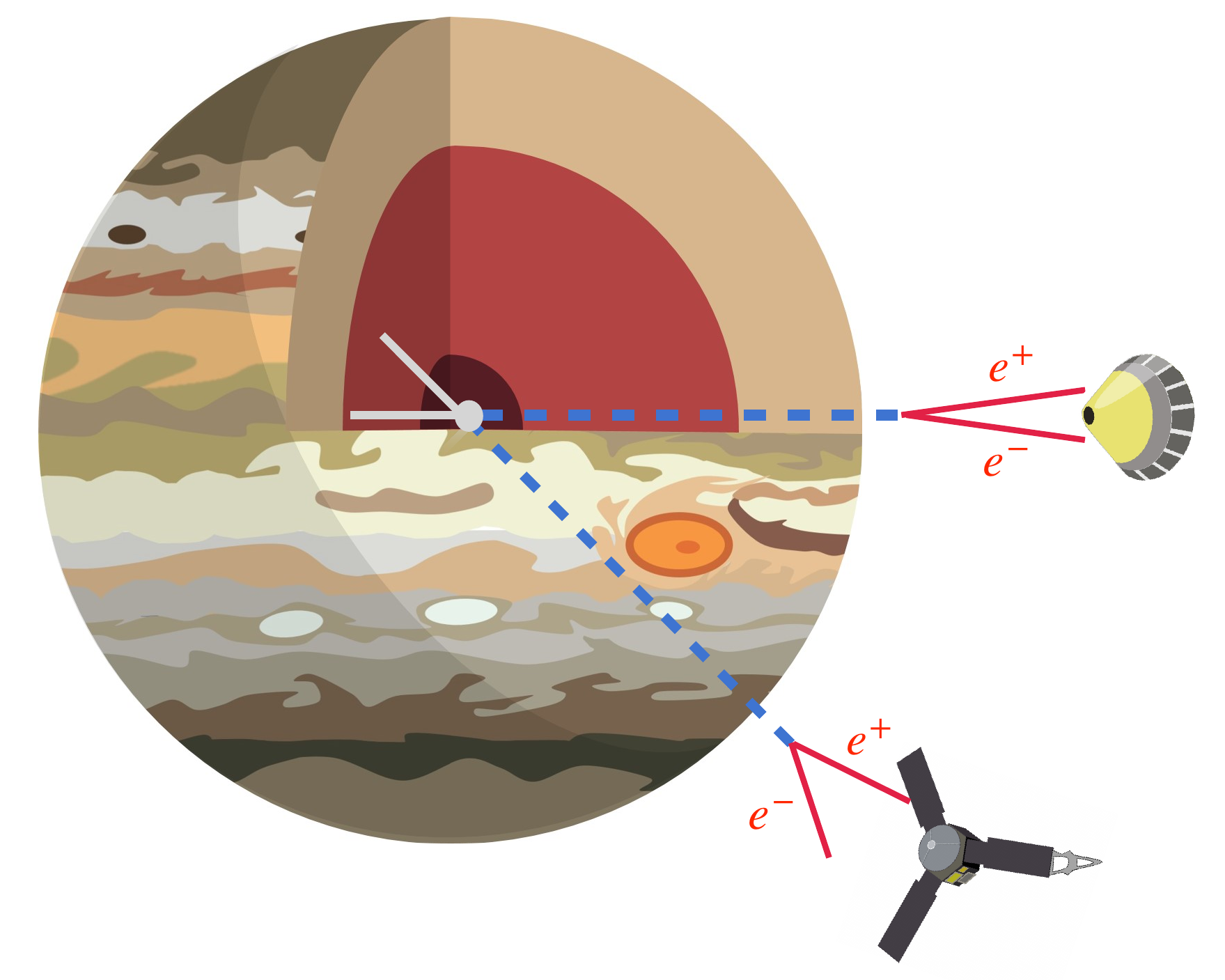}
\caption{A schematic illustration of the process. Captured DM particles (solid white) annihilate into a pair of dark mediators (dashed blue), which decay into $e^+e^-$ (red solid) outside Jupiter. The energetic electrons or positrons could be measured by the Jupiter missions. }
\label{fig:scheme}
\end{figure}

Here we propose a new search for a similar dark sector model but with a different final state: DM annihilating into dark mediators, which decay into a pair of electron $e^-$ and positron $e^+$ outside Jupiter, as depicted in Fig.~\ref{fig:scheme}. Such decay channels could be present and important if long-lived dark mediators couple to standard model fermions. 
One type of the \textit{in situ} measurements, which have been implemented by some Jupiter missions, such as the Galileo probe and the Juno mission, is to profile and measure fluxes of relativistic electrons in the Jovian magnetosphere. For the Galileo probe, its Energetic Particles Investigation (EPI) instrument uses two totally-depleted, circular silicon surface barrier detectors. It made omnidirectional measurements of energetic particle (electrons, protons, $\alpha$-particles, and heavy ions) population in the innermost regions of the Jovian magnetosphere~\cite{fischer1996high}. 
For the Juno mission, the Radiation Monitoring (RM) investigation analyzes the noise signatures from penetrating radiation in the images of Juno's cameras and science instruments~\cite{becker2017juno}. Some Juno instruments, such as the Stellar Reference Unit (SRU), could operate as a star camera collecting sky images by the silicon charge coupled device (CCD) focal plane array. The CCD could register impacts by penetrating charged particles as noise signals, within a cluster of pixels around each hit. These counts could be used to infer the electron fluxes with energy $\gtrsim 10$ MeV at different locations in the radiation belts~\cite{becker2017observations}. While the Jovian electrons as the background of searching for new flux sources are still to be fully understood, the observed fluxes would allow us to set conservative upper bounds on the maximum electron flux induced by dark mediator decays. This could be translated into constraints on un- or under-explored regions of parameter space in the dark sector models. 

Our study serves as a proof of concept and an initial step to explore the full potential of the large datasets from the Jupiter missions to search for new physics beyond the standard model. So far the only other application is to use the Jovian magnetic profile to constrain photon mass~\cite{Davis:1975mn} and dark photon kinetically mixed with the photon~\cite{Marocco:2021dku}, both of which could induce modifications to the magnetic field. Yet Jupiter missions collect much richer information beyond the magnetic field. For example, the relativistic electron flux datasets, the focus of our paper, have not been applied to search for DM and dark mediators before.

The paper is organized as follows. In Section~\ref{sec:capture}, we review the key formalism to compute DM captured by Jupiter. In Section~\ref{sec:model}, we discuss the dark sector scenario that could be probed by the proposed search. In Section~\ref{sec:flux}, we study the motions of electrons trapped in the Jovian magnetic fields and show how to compute the observables related to the electron flux under the approximation of dipole magnetic field and ignoring electron loss effects. In Section~\ref{sec:loss}, we take into account of electron losses, i.e., due to irregular magnetic field lines deviating from the dipole approximation, and classify three possible electron trapping scenarios. In Section~\ref{sec:constraints}, we apply the datasets available to constrain the parameter space of the DM and dark mediator model in interest. We conclude and outline several future directions in Section~\ref{sec:con}.

\section{Dark Matter Capture in Jupiter}
\label{sec:capture}

In this section, we will review the formalism to compute the rate of DM capture in Jupiter. We follow the discussions in Refs.~\cite{Gould:1987ir,Gould:1987ww,1990ApJ...356..302G, Kouvaris:2010vv, Bramante:2017xlb, Dasgupta:2019juq, Ilie:2020vec, Leane:2021ihh, Leane:2021tjj}. We only present the main results and point interested readers to the references above for more details. 

During the capture processes, DM particles, $\chi$'s, from the galactic halo go through one or multiple scattering with the matter in Jupiter and decelerate. It is considered to be captured once DM's relative velocity falls below the escape velocity of Jupiter. For DM with mass $m_\chi$ around the GeV scale, substantial momentum exchange via DM-nucleon elastic scattering is possible as the masses of relevant particles are comparable. Jupiter's stopping power could be described by its optical depth, $\tau_J$:
\begin{equation}
\tau_J = \frac{3}{2}\frac{\sigma_{\chi n}}{\sigma_{\rm sat}}~,~\sigma_{\rm sat}\simeq \frac{\pi R_J^2}{N_{n,J}} \simeq  10^{-34}~\text{cm}^2,
\end{equation}
where $\sigma_{\chi n}$ is the DM-nucleon scattering cross section, which we take to be velocity-independent; $\sigma_{\rm sat}$ is the cross section that saturates the geometric limit; $R_J \simeq 7 \times 10^4$ km is the Jupiter radius; and $N_{n,J}$ is the total number of Jovian nucleons. After one or multiple scattering with nucleons, the DM's velocity becomes comparable or smaller than Jupiter's escape velocity $v_J(R_J) =\sqrt{2 G M_J/R_J} \simeq 59.5$~km/s ($M_J = 1.9 \times 10^{27}$ kg is the Jupiter's mass). In this work, we are interested in the small coupling case, namely the the optically thin limit where $\tau_J \ll 1$ and the capture is achieved in one scattering.\footnote{The approximate multiple scattering formula was first proposed in~\cite{Bramante:2017xlb} and was improved in~\cite{Dasgupta:2019juq, Ilie:2020vec}. However, for Jupiter, the $\sigma_{\chi n}$ needed for multiscattering is much larger than the limit our method can probe. Note that the single scattering limit of the multiscattering formula in~\cite{Bramante:2017xlb} does not match the standard single scattering formula first derived in~\cite{Gould:1987ir,Gould:1987ww}.} In the optically thin limit, we follow the discussions in~\cite{Gould:1987ir,Gould:1987ww}. Assuming a Maxwell-Boltzmann DM velocity distribution and elastic DM scattering all with hydrogen atoms, the capture rate from single scattering could be simplified as:
\begin{equation}
C_1=\sqrt{\frac{8\pi}{3}} \frac{n_\chi \tau_J R_J^2}{\bar{v}_\chi }\int_0^{R_J} \frac{4\pi r^2 n_n(r)}{N_{n,J}}  v_J^2(r)\bigg(1-\frac{1-e^{-A(r)^2}}{A(r)^2}\bigg) X[A(r)] dr~,
\label{eq:capture1}
\end{equation}
where $n_{\chi}$ is the local DM number density (we take the local DM energy density to be 0.4 GeV/cm$^3$~\cite{Sivertsson:2017rkp, Buch:2018qdr, 2020MNRAS.495.4828G, 2020A&A...643A..75S, 2021A&A...653A..86W, 2022MNRAS.511.1977S}), $\bar{v}_\chi \simeq$ 267.2 km/s is DM velocity dispersion~\cite{Freese:2012xd}, $v_J(r)$ is the escape velocity at a distance $r$ from Jupiter's center and is $\lesssim v_J(R_J)$, $m_n$ is the nucleon mass, $n_n(r)$ is the nucleon number density at the distance $r$, $A(r)^2\equiv 6 v_J(r)^2 m_n m_{\chi}/[\bar{v}_\chi^2 (m_n-m_{\chi})^2]$, and $X[A(r)]\in [0.37-0.75]$ is the suppression factor due to the relative motion between Jupiter and the DM halo~\cite{Gould:1987ww, Garani:2021feo}.

Since Jupiter's interior density profile is not completely known, we adopt an approximation by solving the Lane-Emden equation of the polytropic model with $n=1$~\cite{Garani:2021feo}. The resulting density is proportional to $R_J \sin(\pi r/R_J)/\pi r$. The numerical result of Eq.~\eqref{eq:capture1} simplifies to:
\begin{equation}
C_1 \gtrsim  0.28 \sqrt{\frac{8\pi}{3}} \frac{n_\chi \tau_J  R_J^2 v_J^2(R_J)}{\bar{v}_\chi}\bigg(1-\frac{1-e^{-A(R_J)^2}}{A(R_J)^2}\bigg)~.
\end{equation}
To get the right hand side above, we fix $X[A(r)]$ to be the lower end 0.37, while the integration of Eq.~\eqref{eq:capture1} is implemented numerically with the profile of the $n=1$ polytropic model, which gives $\approx 0.7 v_J^2(R_J) \bigg(1-\frac{1-e^{-A(R_J)^2}}{A(R_J)^2}\bigg)$. We use the lower value of $C_1$ for the capture rate throughout the rest of the paper.

DM particles trapped in the planet lose their kinetic energies and accumulate around the center of Jupiter. The total number of trapped DM particles, $N(t)$, evolves as
\begin{equation}
\frac{d N(t)}{d t} = C_1 - \frac{\langle \sigma_{\rm ann} v\rangle}{V_{\rm ann}}N(t)^2~,
\end{equation}
where the second term on the r.h.s. is the rate of DM annihilation, which depletes DM particles; $\langle \sigma_{\rm ann} v\rangle$ and $V_{\rm ann}$ are the thermally averaged DM annihilation cross section and effective volume respectively. The system reaches an equilibrium with a maximum number of DM particles, $N_{\rm max} = \sqrt{C_1 V_{\rm ann}/ \langle \sigma_{\rm ann} v\rangle}$, after a time scale $t_{\rm eq} = (C_1 \langle \sigma_{\rm ann} v\rangle /V_{\rm ann})^{-1/2}$, given $V_{\rm ann}< R_J^3$. As DM eventually thermalizes inside Jupiter, it settles within a length scale of $r_{\rm ann}=\sqrt{9T_J/(4\pi G\rho_J m_{\chi})}$~\cite{Bramante:2017xlb}, where $T_J$ and $\rho_J$ are the characteristic temperature and energy density of the planet's interior region, respectively. Taking the characteristic core temperature of Jupiter as $1.5\times 10^4$~K~\cite{2007ess..book.....M} and the maximum Jupiter core density estimated to be around $2\times 10^4$~kg~m$^{-3}$~\cite{ni2018empirical}, we find that 
\begin{equation}
r_{\rm ann} \simeq 0.1 \, R_J \; \sqrt{\frac{T_{J,{\rm core}}}{1.5\times 10^4~\text{K}}} \sqrt{\frac{1~\text{GeV}}{m_\chi}} \sqrt{\frac{2\times 10^4~\text{kg}~\text{m}^{-3}}{\rho_{J,{\rm core}}}} \, ,
\end{equation}
which is compatible with the typical size of Jupiter's core~\cite{ni2018empirical}. The original discussion of DM evaporation could be found in~\cite{1990ApJ...356..302G}. In particular, the DM evaporation rate is sensitive to its exponential tail originated from either kinematic or thermal distribution even if most DM particles are trapped in a small region~\cite{Garani:2021feo}. For DM lighter than 1 GeV, it could evaporate away before annihilation happens~\cite{Garani:2021feo}. The evaporation boundary varies with $\sigma_{\chi n}$ and Jupiter density profile.

Given the upper limit on the annihilation cross section, $\langle \sigma_{\rm ann}v\rangle \lesssim 5.1 \times 10^{-27} (m_\chi/{\rm GeV})$~cm$^3$~s$^{-1}$ for $m_\chi \geq 1$ GeV from Planck measurements~\cite{Leane:2018kjk}, the time scale $t_{\rm ann}$ for 1 GeV DM is:
\begin{equation}
t_{\rm eq} \simeq 10^{16}~\text{s} \, \sqrt{\frac{5 \times 10^{-27}~\text{cm}^3~\text{s}^{-1}} {\langle\sigma_{\rm ann }v\rangle}} \sqrt{\frac{ 10^{-38}~\text{cm}^2}{\sigma_{\chi,n}}} \, ,
\label{eq:teq}
\end{equation}
which is shorter than the age of Jupiter $t_J \simeq 1.5\times 10^{17}$~s. This also implies that for small $\sigma_{\chi n}$, the equilibrium between DM capture and annihilation may not be reached, depending on the DM mass. In the case when $t_{\rm eq} \gg t_J$, the accumulated number of DM particles will be suppressed by a factor of $t_J/t_{\rm eq}$. In our following computations, we assume the equilibrium is reached and we will revisit this point later at the end of Sec.~\ref{sec:constraints} when discussing constraints from Jupiter missions. 
The total annihilation rate is given by, when the equilibrium is reached, 
\begin{equation}
\Gamma_{\rm ann} \equiv \frac{\langle \sigma_{\rm ann} v\rangle}{2 V_{\rm ann}}N^2~\xrightarrow[ t\gg t_{\rm eq}]{} ~\frac{C_1}{2}~.
\end{equation}
Since $C_1 \propto \sigma_{\chi n}$, $\Gamma_{\rm ann} \propto \sigma_{\chi n}$. 

For readers' convenience, we collect the most important notations throughout the paper and their meanings in Table~\ref{tab:notations}.
\begin{table}[h!]
\begin{center}
\begin{tabular}{ c|l } 
 \hline
 $\chi$ & DM particle \\
 $m_\chi$ & DM mass \\
 $\sigma_{\chi n}$ & DM-nucleon scattering cross section \\
 $\Gamma_{\rm ann}$ & DM annihilation rate  \\
  $\xi$ & dark mediator particle \\
 $\Gamma_D$ & decay width of the mediator \\
 $\gamma$ & mediator boost factor \\ 
$\prod$ BR  & Branching fraction of $e^{\pm}$ from DM annihilations \\
$\equiv$ BR($2\chi\to 2 \xi$) $\times$ BR($\xi\to e^+e^-$) & \\
 $\rho_D$ & mediator decay rate density:  \\
 & decays per unit volume per unit time \\
 $r$ & the radius distance from the Jupiter center \\
  $B$ & magnetic field magnitude\\
 $L$ & McIlwain parameter of the magnetic field lines\\
 $\theta_p$ & geomagnetic latitude $\theta_p$ \\
 $\alpha$ & pitch angle \\
 $\alpha_{\rm eq}$ & equatorial pitch angle \\ 
 $E$ & energy \\ 
  $\bar{I}$ & averaged injection rate of $e^\pm$ over their trajectories \\ 
 $f$ & the electron phase space distribution \\ 
 $\tau_E$ & time scale of electron energy loss \\
 $\tau_y$ & time scale of pitch angle variation \\  
 $\tau_{\rm loss}$ & time scale of electron loss \\
 $J(L, \theta_p)$ & omnidirectional number flux of relativistic $e^{\pm}$  \\
 & integrated over the measured energy range  \\
 &in the $L$-shell at $\theta_p$ \\
 $F$ & geometric factor: effective collecting area \\
 $K (\tilde{K})$ & observed (predicted) count rates:  \\
 & number of electron hits recorded per unit time \\
  $J_{\rm inf} (\tilde{J}_{\rm inf})$ & observed (predicted) omnidirectional fluxes  \\
  & inferred from count rates \\ 
 \hline
\end{tabular}
\end{center}
\caption{Important notations and their meanings.}
\label{tab:notations}
\end{table}

\section{Dark Sector with a Long-lived Mediator}
\label{sec:model}

In this section, we present and discuss the DM scenario that could be probed by the Jupiter \textit{in situ} measurements on the flux of energetic electrons. 

In the scenario being considered, captured DM particles annihilate into a pair of long-lived mediators, $\xi$'s, with decay lengths comparable to or even longer than Jupiter's radius $R_J \simeq 7\times 10^4 $~km, which means that a significant fraction of $\xi$ decay outside the planet. This is possible if $\xi$ is feebly coupled to and decays to the standard model. Possible candidates of $\xi$ include either dark photon kinetically mixed with the standard model photon~\cite{Kobzarev:1966qya, Okun:1982xi, Galison:1983pa, Holdom:1985ag, Holdom:1986eq} or heavy axion-like particles that couple to leptons (e.g, MeV-10 GeV axion accompanying DM and coupling to leptons could arise naturally in low-scale supersymmetry breaking model~\cite{Nomura:2008ru, Ibe:2009dx, Mardon:2009gw, Fan:2010is}). Decay products with energy $\sim \mathcal{O}(m_{\chi})$ will be released to the planet's radiation belt. In the following, we only focus on the $2\chi \to 2\xi$ annihilation process, ignoring all other annihilation channels. 
In principle, the interaction responsible for capture of DM by nucleons could also lead to DM annihilation into standard model particles. Yet this may not need to be the dominant annihilation channel. For example, one simplest effective operator that gives rise to spin-dependent DM-nucleon scattering is 
\begin{equation}
\frac{g_{\chi}g_{q}}{\Lambda^2} \left(\bar{\chi}\gamma^\mu \gamma^5 \chi\right) \left(\bar{q} \gamma_\mu \gamma^5 q\right) \, ,
\end{equation}
with DM $\chi$ being a Dirac fermion and $q$ the standard model quarks. This could be generated via integrating out an axial-vector with a mass about the scale $\Lambda$ and coupling $g_\chi$ ($g_q$) to DM (quark). The resulting spin-dependent DM-nucleon scattering cross section is~\cite{Fan:2010gt, Abdallah:2015ter}
\begin{equation}
\sigma_{\chi n} \approx 3.8 \times 10^{-39} \, {\rm cm}^2 \, \left(\frac{\mu_{\chi n}}{\rm GeV}\right)^2 \, \left(\frac{g_\chi g_q}{ 10^{-3}}\right)^2  \, \left(\frac{10 \, {\rm GeV}}{\Lambda}\right)^4 \, ,
\end{equation}
where $\mu_{\chi n}$ is the reduced mass of DM and nucleon and we take $g_q$ to be the same for $u, d$ and $s$ quarks. The benchmark value here is chosen to satisfy current direct detection bound for $m_\chi$ of order a few GeV and below, as well as evade collider constraints. 
The cross section of annihilation $\bar{\chi}\chi \to \bar{q}q$ through this operator is suppressed by both the small $(g_\chi g_q)^2$ and $m_\chi^2 m_q^2/\Lambda^4$ and could be significantly below the CMB bound when we consider GeV-scale DM. On the other hand, we would consider that the fermionic DM is charged under a dark $U(1)$ gauge symmetry and annihilates into dark photons $A^\prime$ via the $\bar{\chi}\chi \to A^\prime A^\prime$ process. The thermally averaged annihilation cross section is
\begin{equation}
\langle \sigma v \rangle (\bar{\chi}\chi \to A^\prime A^\prime) \approx 2 \times 10^{-27} {\rm cm}^{3} \, {\rm s}^{-1} \left(\frac{\alpha_D}{5 \times 10^{-6}}\right)^2 \, \left(\frac{\rm GeV}{m_\chi}\right)^2 \, ,
\end{equation}
where we assume $m_{A^\prime} \ll m_\chi$. This annihilation channel could dominate over the annihilation into the standard model and saturate the CMB constraint simultaneously. Note that the overall small annihilation cross section could lead to too much thermal relic abundance but could be diluted in a non-thermal scenario, e.g., with early matter domination and another reheating after the inflationary reheating but before BBN. Another possible mechanism that could achieve a right thermal relic is the interplay of co-annihilation and co-scattering processes between the DM, its nearly-degenerate partner, and the dark photon~\cite{Cheng:2018vaj}. For the rest of the discussion, we will be agnostic of the relic abundance mechanism.

\begin{figure}[h!]
\centering
\includegraphics[width=9.5 cm]{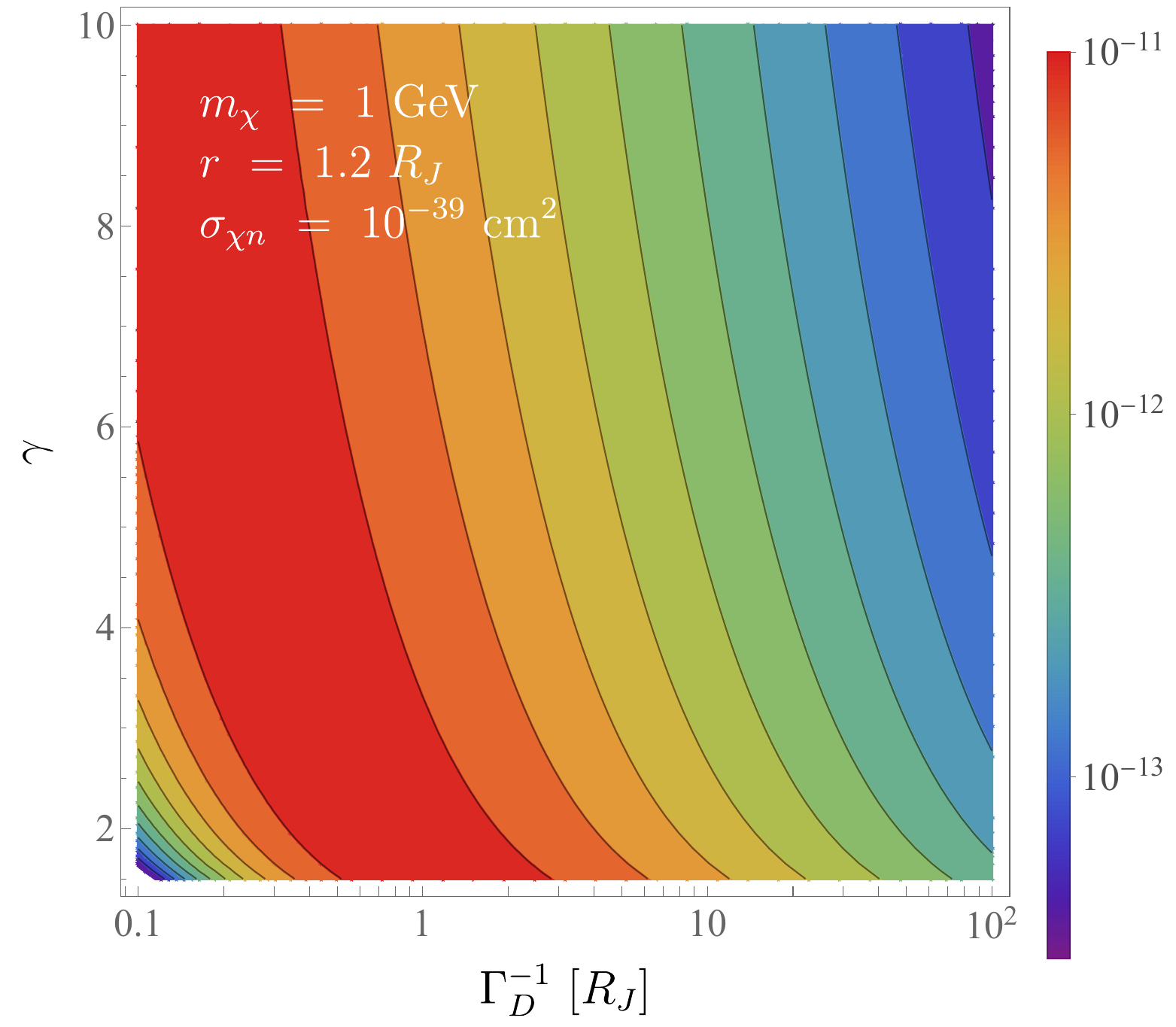}
\caption{Mediator decay density rate, $\rho_D$, at $r=1.2 R_J$ in unit of cm$^{-3}$~s$^{-1}$, as a function of the mediator lifetime $\Gamma_D^{-1}$ in unit of $R_J$, and the boost factor $\gamma$. }
\label{fig:rate_density}
\end{figure}

The scattering of $\xi$ with the Jovian matter before its escapes could be ignored due to its very weak interactions. 
From the discussion in the last section, the annihilation region would mostly be in the core of Jupiter. It is a reasonable approximation that all DM annihilate, and all $\xi$'s are produced at the center of Jupiter. 
The number of mediator decays per second per unit volume, the mediator decay rate density at a distance $r$ from the Jupiter center takes the simple form:
\begin{equation}
\rho_{D}(r) = \frac{2\Gamma_{\rm ann}}{4\pi r^2} \frac{\Gamma_{\rm D}}{\gamma \beta} e^{-\frac{\Gamma_{\rm D} r}{ \gamma\beta}}~,
\label{eq:decaydensity}
\end{equation}
where $\Gamma_{\rm D}$ is the decay width of the mediator; the mediators from DM annihilations are boosted by a factor of $\gamma\simeq m_\chi/m_{\xi}$ with $m_{\xi}$ the mediator mass; and the velocity of the mediator is given by $\beta = \sqrt{1-\gamma^{-2}}$. The factor of 2 in the numerator is due to the fact that each annihilation produces two mediators. Fig.~\ref{fig:rate_density} depicts $\rho_D$ as a function of $\gamma$ and $\Gamma_{\rm D}^{-1}$. From the figure, one could see that at a given $r$ close to the Jupiter surface, $\rho_D$ is maximized when $\Gamma_D^{-1}\times \gamma \sim R_J$. For $\Gamma_D^{-1} \times \gamma \ll R_J$, $\rho_D$ is suppressed by the exponential factor in Eq. \eqref{eq:decaydensity} since most decays happen inside Jupiter. For $\Gamma_D^{-1} \times \gamma \gg R_J$, $\rho_D$ is suppressed by the small $\Gamma_D$ factor in Eq. \eqref{eq:decaydensity} since not many decays happen in the innermost radiation belt.

One of the most interesting decay channels of a sub-GeV mediator is the decay to energetic electrons and positrons, $e^\pm$. For example, a dark photon, $A^\prime$, decays to $e^+e^-$ exclusively if its mass $m_{A^\prime}$ is below the $2m_\mu$ threshold but above the $2m_e$ threshold. Above the $2m_\mu$ threshold, the decay products becomes a mixture of $e^+e^-$, $\mu^+\mu^-$ and $\pi^+\pi^-$, depending on $m_{A^\prime}$~\cite{Pospelov:2007mp, Arkani-Hamed:2008kxc, Reece:2009un, Bjorken:2009mm, Ilten:2018crw}. Most non-electron final states will cascade to a pair of softer electrons and a few neutrinos, leaving almost no other visible particles. The discussion of decay products above also applies to axion-like particles coupling to standard model fermions. We will only consider the channel of dark mediators decaying into $e^+e^-$ in the following discussion. The density rate of $e^+e^-$ final state is then
\begin{equation}
\rho_D(r) \prod {\rm BR}, \quad  \prod {\rm BR} \equiv {\rm BR}(2\chi\to 2 \xi) \times {\rm BR}(\xi\to e^+e^-) \, ,
\end{equation}
where $\prod {\rm BR}$ is the product of the branching fractions of DM annihilating into dark mediators and $\xi$ decaying to $e^{\pm}$. 

Requiring the mediator's decay length comparable to $R_J$ allows us to explore the parameter space that is difficult to be probed otherwise. Take the dark photon as an example. 
To have a decay length of order $R_J$, its decay width satisfies
\begin{equation}
\Gamma_{\rm D} (A^\prime \to e^+e^-)=\frac{\alpha}{3}\epsilon^2 m_{A^\prime} \sqrt{1-\frac{4 m_e^2}{m_{A^\prime}^2}} \left(1+\frac{2m_e^2}{m_{A^\prime}^2}\right)\sim R_J^{-1}~\Rightarrow~\epsilon^2 \sim 10^{-20}\times\bigg(\frac{0.1~\text{GeV}}{m_{A^\prime}}\bigg)~,
\end{equation} 
where $\epsilon$ is the kinetic mixing parameter, $m_{A^\prime}$ is the dark photon mass; $m_e$ is the electron mass; and $\alpha \simeq 1/137$ is the standard model fine structure constant. 
Such a small mixing $\epsilon$ does not introduce significant interactions between the mediator and the standard model, leaving $A^\prime$ above 0.1 GeV elusive in terrestrial and cosmological searches~\cite{Alexander:2016aln, Fabbrichesi:2020wbt}. We also stress that the cross section of $A^\prime$ mediated DM-electron scattering is suppressed by the tiny $\epsilon^2$ and is irrelevant for the current sub-GeV DM-electron scattering searches. Similarly, the contribution to DM capture from DM-nucleon scattering processes mediated by $A^\prime$ is negligible. In other words, DM capture (DM-nucleon scattering) and annihilation (DM annihilates into dark mediators) are two independent processes in our scenario.

Electrons from $\xi$ decays are injected into the Jovian magnetosphere and can be trapped in the strong magnetic field for a long time. The phase space evolution of the high-energy electrons will be detailed in the next two sections. The long time scale of the trapped electrons compensates for the low $\rho_D$ when the DM-nucleon scattering cross section $\sigma_{\chi n}$ is small. In addition, the hard electron spectrum originating from the mass scale of DM is distinctive from the soft astrophysical background~\cite{sicard2011jose}. These features make several \textit{in situ} measurements of Jupiter missions sensitive to this class of models.

\section{Relativistic $e^\pm$ Fluxes in the Innermost Radiation Belt: Dipole Approximation}
\label{sec:flux}

In this section, we describe the motion of $e^{\pm}$ from $\xi$ decaying inside the Jovian magnetosphere and derive the basic formula for the resulting flux of charged particles. We will first work with the approximation that the Jovian magnetic field could be described as a dipole. Note that \textit{in situ} measurements do not distinguish between electrons and positrons. The electron in the following discussions includes both $e^-$ and $e^+$.

As an approximate magnetic dipole, the Jovian magnetic field is significantly stronger near the polar regions. Charged particles traveling along the magnetic flux tube will then be reflected due to the magnetic mirror effects and eventually trapped inside the radiation belts. It is thus convenient to introduce the McIlwain $L$ parameter~\cite{mcilwain1961coordinates} to describe the magnetic field lines. Here, an $L$-shell could be understood as a collection of magnetic flux tubes within which charged particles drift through. It is displaced from the Jupiter center by an amount $L R_J$ in the magnetic equatorial plane under the dipole approximation. As the $B$ field only changes slowly along the field lines, the overall volume of a flux tube in an $L$-shell $V_L=\int_L dV$ can be approximated as $\int_L dA \cdot dS$, in which $S$ stands for the length along the field line and the cross sectional area $A \propto |B|^{-1}$ since the magnetic flux is a constant. Therefore, the overall rate of electrons injected from the mediator decays into each flux tube reads $2\int_L \rho_D~dA  \cdot dS$, with the factor of 2 taking both $e^-$ and $e^+$ into account. 

\begin{figure}[h!]
\centering
\includegraphics[width=10cm]{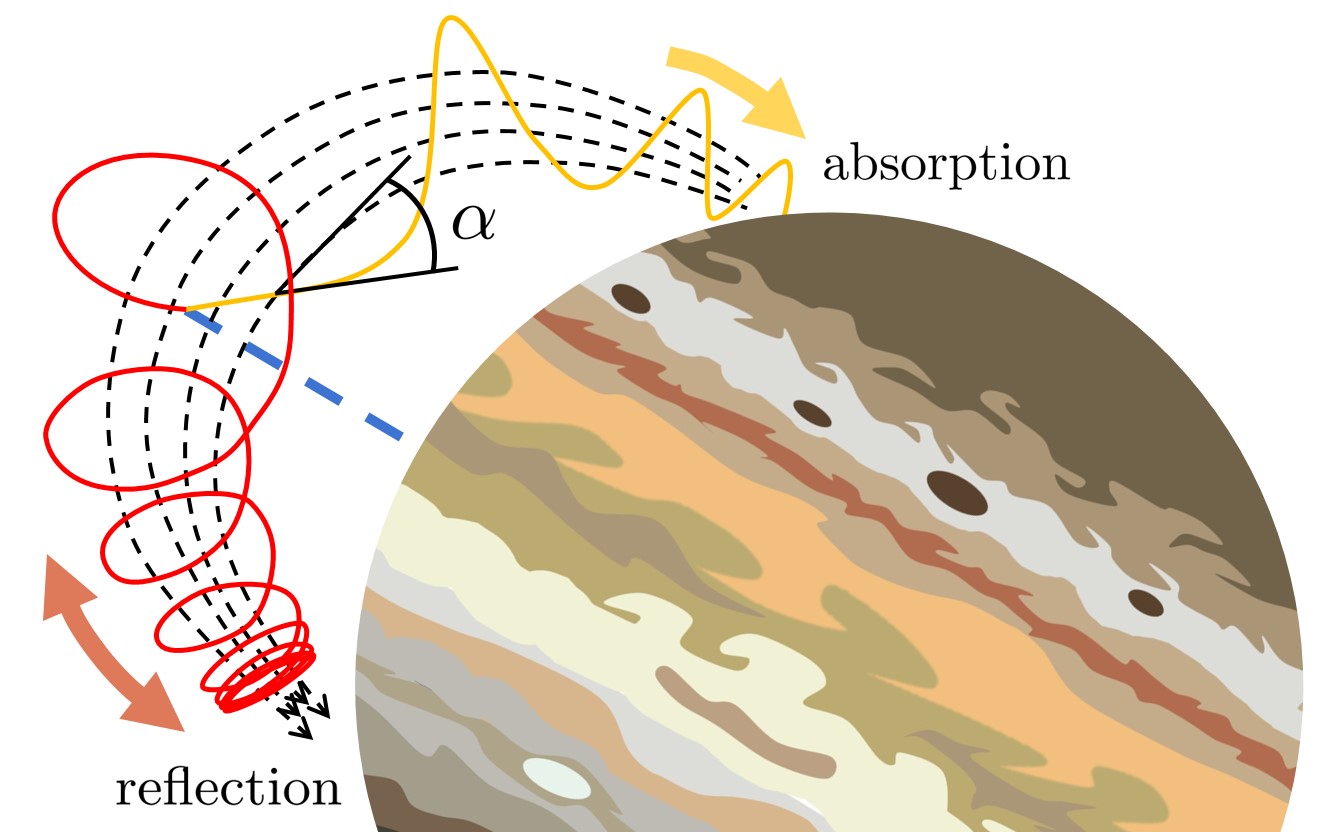}
\caption{A cartoon of the electron motions in the Jovian magnetic fields (field lines are denoted by the black dashed curves). The dark mediator will travel outside Jupiter (blue dashed line) and decay to a pair of $e^\pm$ (which we will refer to as electrons). The electrons gyrate around the magnetic flux tube and travel along the field lines. For the electron with a large initial pitch angle $\alpha$, as $B$ increases along its trajectory, its $\alpha$ eventually reaches $\pi/2$ around the mirror point and gets reflected, as demonstrated by the red trajectory. In contrast, the electron with a small initial $\alpha$ will hit and be absorbed by the atmosphere before reaching the mirror point, as shown by the yellow trajectory.}
\label{fig:emotion}
\end{figure}

In addition to energy $E$ and $L$, an extra kinematic variable is needed to describe the phase space of the trapped charged particles. It is taken to be the equatorial pitch angle $\alpha_{\rm eq}\equiv \arcsin (|p_\perp|/|p|)|_{\rm eq}$, which is the pitch angle between the momentum of the particle and the magnetic field at the magnetic equator ($p_\perp$ is the component of the momentum perpendicular to the magnetic field line). As the particle's magnetic moment $M=|p_\perp|^2/(2mB)$ is adiabatically invariant along its trajectory, we have the following relation for each $L$-shell:
\begin{equation}
 \alpha_{\rm eq} = \arcsin \bigg( \sqrt{\frac{B(0)}{B(\theta_p)}}\sin \alpha(\theta_p) \bigg)~,
 \label{eq:alpharelation}
\end{equation}
where $\alpha(\theta_p)$ and $B(\theta_p)$ are the pitch angle and magnetic field at the geomagnetic latitude angle $\theta_p$, while $B(0)$ is the equatorial magnetic field. As the particle travels to higher $|\theta_p|$, $\alpha$ increases with $B(\theta_p)$ and eventually the electron gets reflected around the two mirror points where $\alpha=\pi/2$. Even with the same energy and $L$-shell value, the trapping time scales and spatial distributions of electrons vary when their $\alpha_{\rm eq}$'s change. Moreover, all particles with $\alpha_{\rm eq} < \alpha_{\rm min}$ will be absorbed by the Jovian atmosphere. Here $\alpha_{\rm min}$ is the minimum equatorial pitch angle that makes $\sin\alpha(\theta_p)=\pi/2$ in Eq.~\eqref{eq:alpharelation} at the $\theta_p$ where $r(L,\theta_p)=R_J$. Electrons with even smaller $\alpha_{\rm eq}$ will have their mirror points inside Jupiter and be absorbed by the atmosphere. Thus it is necessary to include the pitch angle distribution when deducing the electron fluxes. We illustrate the two possible types of electron motions ($\alpha_{\rm eq}>\alpha_{\rm min}$ and vice versa) in Fig.~\ref{fig:emotion}.

On top of the gyration around the magnetic flux tube and the bounce between mirror points along the magnetic field lines, there is a third motion, i.e. the drift in the longitudinal direction around the planet~\cite{schulz2012particle}. The drift stems from the gradient of the magnetic field and has a much longer timescale compared to the gyration and bounce. Such a motion doesn't affect the electron flux as long as the magnetic field is azimuthally symmetric. Its effect when the azimuthal symmetry is broken will be discussed in Section~\ref{sec:loss}.

The injected electrons from $\xi$ decays serve as an extra source term of energetic particles in the radiation belt. We denote the phase space distribution of the $e^\pm$ injection rate as $I_{e^-}=I_{e^+}\equiv I$. At each local point, it is a function of $r, L, E, \sin \alpha$\footnote{Since $r=L \cos^2 \theta_p$ in a dipole field, one could also think of $I$ as a function of $\theta_p, L, E, \sin \alpha$ in this case.} and is related to the mediator decay rate density $\rho_D$, which only depends on $r$, as:
\begin{equation}
2  \int dE\, E^2 \int d\phi \int  d \alpha \, \sin\alpha  \, I(r,L,E,\sin \alpha)   = \rho_D(r)~,
\label{eq:rhoD}
\end{equation}
where $\phi$ is the gyration angle. In practice, the injected electron rate distribution $I$ is determined by Monte-Carlo simulations for a given DM model. We will describe the simulations in App.~\ref{app:simulations}.

\begin{figure}[h]
\centering
\includegraphics[width=7.8 cm]{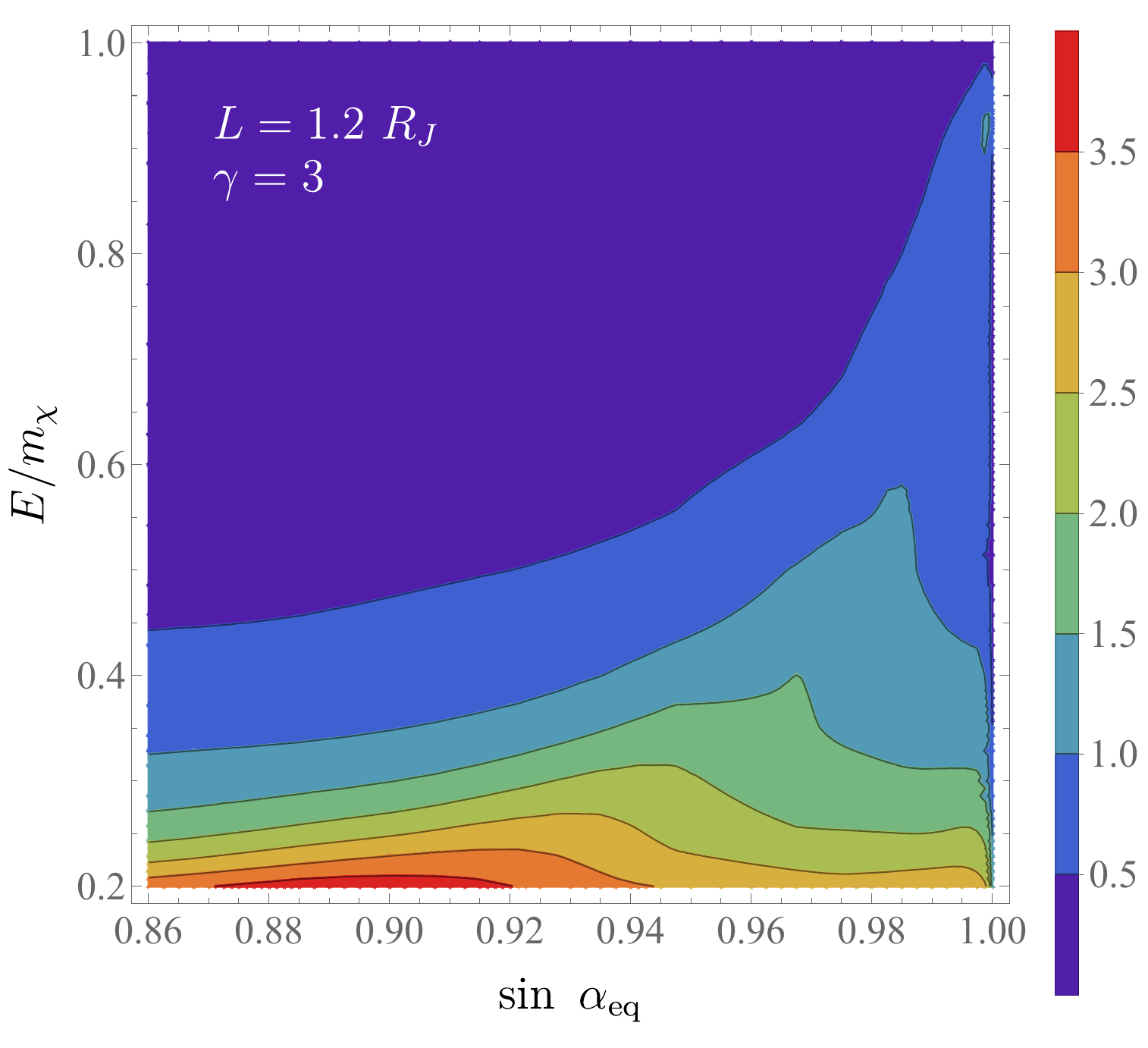}
\includegraphics[width=7.5 cm]{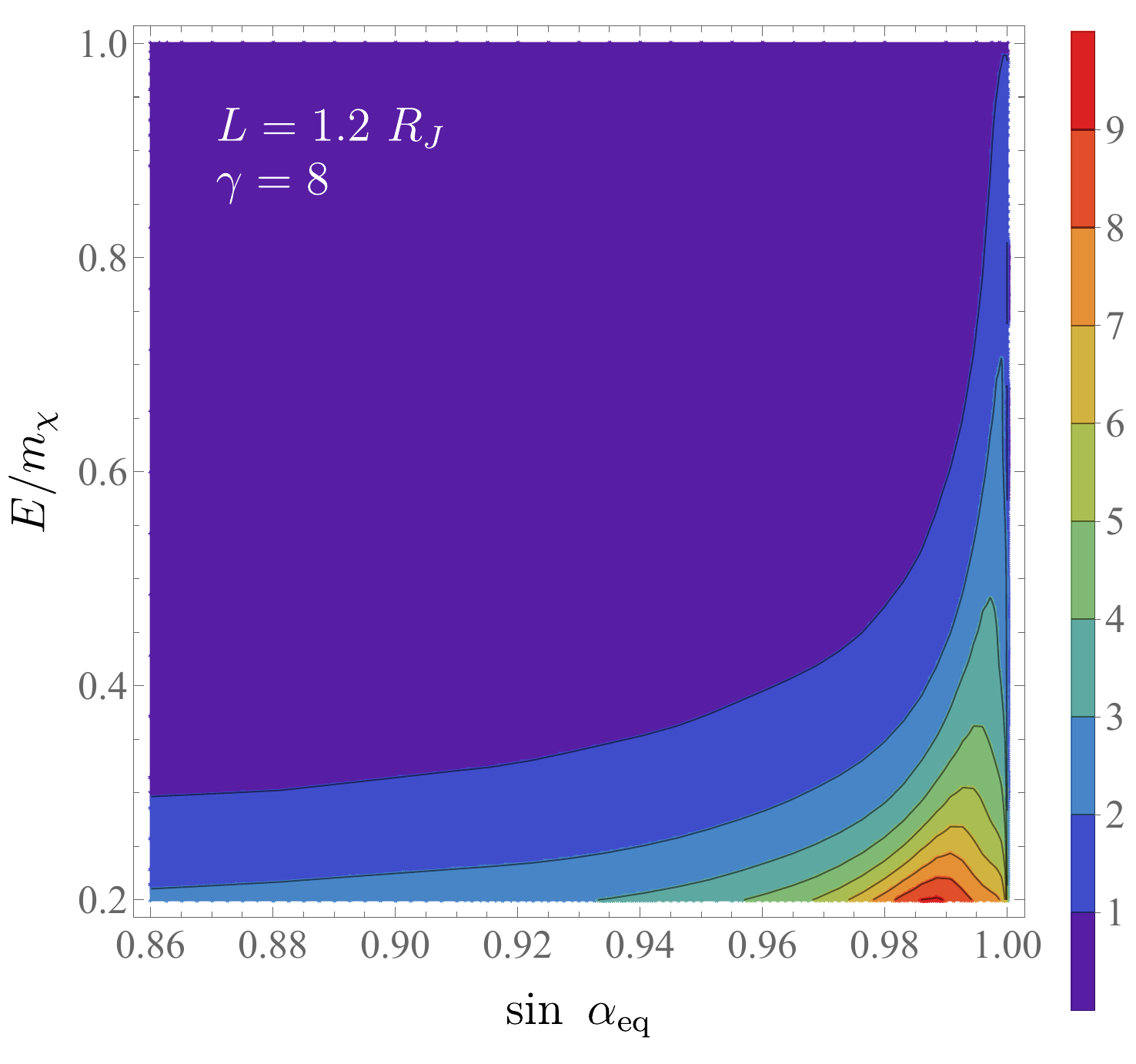}
\caption{The trajectory-averaged injection rate $\bar{I}$, as a function of $E$ and $\alpha_{\rm eq}$ for different model parameters. The presented values are in the unit of $m_\chi^{-3} \rho_D$ and thus dimensionless. In both cases, contributions from the large-$\alpha_{\rm eq}$ regime are significant due to $\gamma$ greater than 1. {\bf LEFT:} The distribution of $\bar{I}$ at $L=1.2$, from a DM model with $\gamma$=3. {\bf RIGHT:} The same as the left, but with $\gamma=8$.}
\label{fig:form1}
\end{figure}

When reaching equilibrium, the diffusion equation of the electron phase space density $f(L, E, \sin \alpha_{\rm eq})$ is~\cite{nenon2018rings}:
\begin{align}
\label{eq:simple}
\frac{d f(L, E,\sin \alpha_{\rm eq})}{d t} &= \frac{\int I dz}{\int dz} \nonumber \\
&- \frac{1}{G}\frac{\partial }{\partial E}\bigg( \frac{d E}{d t}Gf \bigg)  - \frac{1}{G}\frac{\partial }{\partial \sin \alpha_{\rm eq}}\bigg( \frac{d \sin \alpha_{\rm eq}}{d t} G f \bigg) \nonumber  \\
& +\rm{loss~terms} ~+~ \rm{diffusion~terms}=0 \, ,
\end{align}
where $z$ is the length of electron trajectories, and the first term $\int I dz/\int dz\equiv  \bar{I}$ is the averaged electron injection rate over their trajectories as the source term. Some benchmark $\bar{I}$ are depicted in Fig.~\ref{fig:form1} as functions of $E$ and $\sin\alpha_{\rm eq}$. The presented values are in the unit of $m_\chi^{-3} \rho_D$ and thus dimensionless. For boosted $\xi$ decays, the energy distribution of injected electrons is a uniform one with $E$ between $(1\pm\sqrt{1-\gamma^{-2}})m_\chi/2$. Consequently, both $I$ and $\bar{I}$ scales as $\sim E^{-2}$ to keep $E^2 I(\bar{I})$ approximately a constant. From Fig.~\ref{fig:form1}, one can also notice that the contribution of large $\alpha_{\rm eq}$ is bigger. This is because the boost makes injected electrons closer to the direction perpendicular to the magnetic field lines. Such an effect becomes more significant when $\gamma$ increases from 3 to 8. 

The factor $G $ in Eq.~\eqref{eq:simple} is the Jacobian from the three action variables defined in~\cite{schulz2012particle} to ($L,E,\sin \alpha_{\rm eq}$).\footnote{Under the dipole approximation and in the relativistic limit, $G$ can be approximated as $E^2 \sin \alpha _{\text{eq}} \left(3.84 \sin \alpha _{\text{eq}}+3.84 \sqrt{\sin \alpha _{\text{eq}}}-16.56\right)$ times a function of $L$~\cite{schulz2012particle}.} 
The two terms in the second line of Eq.~\eqref{eq:simple} stand for the friction effects (e.g., energy loss and pitch angle change with time), each associated with a characteristic time scale. Notice that with friction terms only, $L$ will remain constant. The loss term describes the removal of electrons due to the deviation from the dipole approximation and hard scattering processes, which will be detailed in Section~\ref{sec:loss}. In a dipole field, the effect of loss terms is insignificant. Diffusion terms in Eq.~\eqref{eq:simple} include radial, energy, and angular diffusion effects from various origins like the interchange instability of plasma~\cite{1987JGR....92..109S} or interactions with the low-frequency plasma waves~\cite{2005JGRA..110.4206G}, which are orders of magnitude smaller than the friction terms within the innermost region ($L\lesssim 1.5$) we are interested in~\cite{nenon2017new}. During the typical trapping time scale of $e^\pm$, they will not affect the solution of $f$ significantly and thus can be safely ignored.

\subsection{Friction Time Scales}
\label{sec:frictionsync}
\begin{figure}[h!]
\centering
\includegraphics[width=7.3 cm]{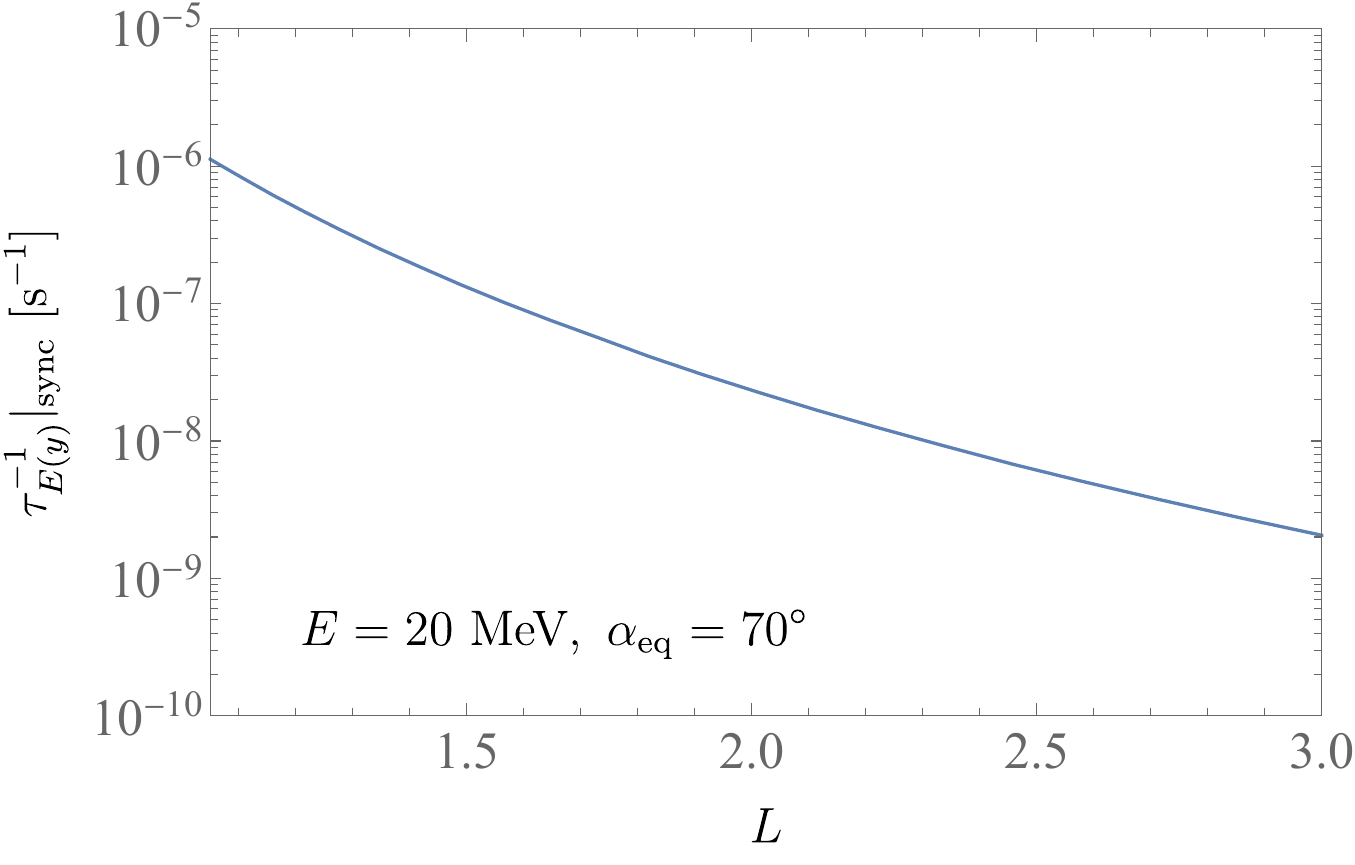}
\includegraphics[width=7 cm]{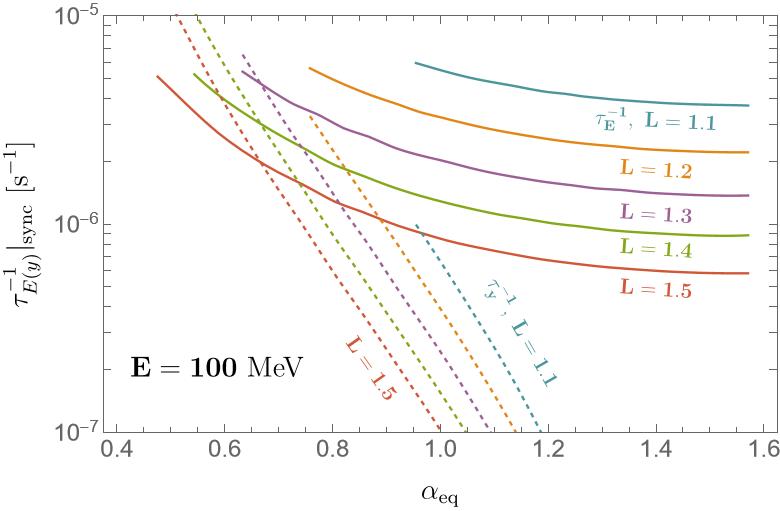}
\caption{Time scales of synchrotron radiation with the magnetic dipole approximation. {\bf LEFT:} The energy loss rate for 20~MeV electrons with $\alpha_{\rm eq}=70^\circ$ as a function of $L$, the result is compatible with the result reported in~\cite{nenon2017new,nenon2018rings}. {\bf RIGHT:} Contours of $\tau_E^{-1}$ (solid) and $\tau_y^{-1}$ (dashed) as functions of $\alpha_{\rm eq}$. The five colors, from top to bottom for $\tau_E^{-1}$ and from right to left for $\tau_y^{-1}$, stand for $L=1.1,~1.2,~1.3,~ 1.4$, and $1.5$, respectively.}
\label{fig:Sync}
\end{figure}

In this section, we will estimate the friction effects and the associated time scales in Eq.~\eqref{eq:simple}. The energy loss rate $dE /dt$ receives various contributions, with the leading one being the synchrotron radiation in our scenario. The energy loss rate from synchrotron radiation is\footnote{The energy loss rates in this section should all be taken as absolute values.} 
\begin{align}
\frac{1}{E}\frac{d E}{d t}\bigg|_{\rm sync}&= \sigma_T \frac{E}{m_e^2}B^2\sin^2 \alpha~ \nonumber \\
&\simeq 6.3\times 10^{-6}  \sin^2 \alpha \bigg(\frac{E}{100~\text {MeV}}\bigg) \bigg(\frac{B}{4~\text{Gauss}}\bigg)^2\, \text{s}^{-1} \,~, 
\label{eq:syn}
\end{align}
where the Thomson cross section $\sigma_T=8\pi \alpha^2/(3 m_e^2)  \approx 6.7\times 10^{-29}$~m$^2$. Since a trapped electron bounces inside the $L$-shell multiple times before its kinetic energy drops significantly, the time scale, $\tau_E$, can be calculated by averaging over the trajectory parametrized by the trajectory length $z$:
\begin{equation}
\tau_E^{-1}|_{\rm sync}\equiv \left\langle\frac{1}{E}\frac{d E}{d t}\bigg|_{\rm sync}\right\rangle=\sigma_T  \frac{E}{m_e^2} \frac{ \int B^2 \sin^2 \alpha \, dz}{\int dz}~.
\label{eq:inversescale1}
\end{equation}
From Eq.~\eqref{eq:syn}, one could see that the characteristic time scale of synchrotron radiation is $\tau_E|_{\rm sync} \sim \mathcal{O}(10^5)$~s for electrons with energies of ${\cal O}(100)$ MeV. Taking the dipole approximation with the intensity from the JRM09 magnetic field model~\cite{connerney2018new}, we present the energy loss rate as a function of $L$ in the left panel of Fig~\ref{fig:Sync}.  As $\tau_E|_{\rm sync} \propto B^{-2}$ and $B \sim L^{-3}$ in a dipole field, $\tau_E^{-1}|_{\rm sync}$ scales as $L^{-6}$.

The synchrotron radiation also alters the electron's pitch angle. The time scale $\tau_y$ of the pitch angle variation is proportional to the energy loss rate~\cite{santos2001modeling}:
\begin{equation}
\tau_y^{-1}|_{\rm sync}\equiv \left\langle \frac{d \sin\alpha_{\rm eq}}{d t}\bigg|_{\rm sync} \right\rangle = \frac{\cos^4 \alpha_{\rm eq}}{\sin^2\alpha_{\rm eq}} \left\langle\frac{1}{E} \frac{d E}{d t}\bigg|_{\rm sync}\right\rangle~.
\label{eq:inversescale2}
\end{equation}
For $\alpha_{\rm eq}\gtrsim \pi/4$, the timescale of pitch angle change $\tau_y$ is longer than $\tau_E$. This could be seen from the right panel of Fig~\ref{fig:Sync}, in which the inverse time scales $\tau_E^{-1}|_{\rm sync}$ and $\tau_y^{-1}|_{\rm sync}$ are plotted. This means that for large enough $\alpha_{\rm eq}>\pi/4$, $\tau_E$ is the more relevant time scale compared to $\tau_y$. Another crucial feature shown in Fig.~\ref{fig:Sync} is that as $\alpha_{\rm eq}$ increases, both the friction rates become smaller, or equivalently, $\tau_E$ and $\tau_y$ becomes longer. From Eq.~\eqref{eq:syn}, \eqref{eq:inversescale1} and \eqref{eq:inversescale2}, one might expect that $\tau_E^{-1}|_{\rm sync}$ and $\tau_y^{-1}|_{\rm sync}$ increases as $\alpha_{\rm eq}$ increases. However, with larger $\alpha_{\rm eq}$, electrons tend to stay near the magnetic equator (in the limit that the initial $\alpha_{\rm eq} = \pi/2$, the electron just stays in the equatorial plane), where the magnetic field is the weakest. Smaller $\alpha_{\rm eq}$ allows the electrons to travel to regions with denser magnetic field lines near the poles. It turns out that $\tau_E^{-1}|_{\rm sync}$ and $\tau_y^{-1}|_{\rm sync}$ are more sensitive to variation in the $B$ field and thus decreases as $\alpha_{\rm eq}$ increases.

An electron could also lose energy due to Coulomb scattering with gas and plasma along the trajectory. Furthermore, frequent hard scatterings transferring electron energy to gas lead to electron absorption, which will then be better described as a loss term instead of a friction term in Eq.~\eqref{eq:simple}. This applies to electrons diving deeply into the Jovian atmosphere. For observations performed in regions where the gas density is sufficiently small~\cite{hinson1998jupiter,bougher2005jupiter}, the energy friction term of Coulomb scattering is significantly smaller than that of the synchrotron radiation and thus could be safely ignored. The electron loss time scale, on the contrary, may affect electrons with $r$ up to $\sim 1.3 R_J$ and will be discussed in Section~\ref{sec:loss}.

Beyond the synchrotron radiation and Coulomb scattering effects, other sources may also introduce extra friction terms, like interaction with grains in the halo ring~\cite{nenon2018rings}. However, for energetic electrons ($E\gtrsim 10$~MeV) within $L\lesssim 1.5$ region, such extra friction terms are all highly suppressed compared to the synchrotron radiation and can also be safely ignored for our purpose~\cite{nenon2018rings}.

\subsection{Electron Flux}

The omnidirectional number flux of ultra-relativistic $e^\pm$ (the number of $e^\pm$ passing through unit cross section per unit time) with energies above a threshold $E_{\rm th}$, in the $L$-shell at the geomagnetic latitude $\theta_p$, could be calculated from the phase space distribution $f$:
\begin{equation}
J(L,\theta_p)|_{E>E_{\rm th}} =  2\int_0^{2\pi} d \phi\int_0^{\frac{\pi}{2}}  \sin \alpha_{\rm eq} \, d\alpha_{\rm eq} \int_{E_{\rm th}}^{+\infty} E^2 \, dE \, f(L,E,\sin\alpha_{\rm eq}) \frac{d A_{\rm eq}}{d A}\frac{(dt/dS)}{(dt/dS)_{\rm eq}} ~,
\label{eq:Jexact}
\end{equation}
where the factor of 2 in front captures the symmetric particle population moving along both directions of the magnetic flux tube. Here we choose to integrate over the equatorial pitch angle. Away from the equatorial plane, we need to include the last two factors in the equation above: $d A_{\rm eq}/d A$ encodes the enhancement of the electron flux as the magnetic flux tube narrows at higher latitudes; $\frac{(dt/dS)}{(dt/dS)_{\rm eq}}$ takes into account that the electron's speed along the magnetic field line varies: $dS/dt = v_\parallel = \cos \alpha$, and the electron spends less time near the equatorial region. 
Using Eq.~\eqref{eq:alpharelation} and $dA \propto B^{-1}$, Eq.~\eqref{eq:Jexact} is further reduced to:
\begin{align}
J(L,\theta_p)|_{E>E_{\rm th}} &=  4\pi  \int E^2 dE \int  \sin \alpha_{\rm eq} d\alpha_{\rm eq}  \, f(L,E,\sin\alpha_{\rm eq}) \frac{B(\theta_p)}{B(0)} \frac{\cos \alpha_{\rm eq}}{\cos \alpha}   \nonumber \\
& =  4\pi \int E^2 dE \int  f(L,E,\sin\alpha_{\rm eq}) \bigg(\sqrt{\frac{B(\theta_p)}{B(0)}}\sin \alpha_{\rm eq}\bigg) \bigg(\sqrt{\frac{B(\theta_p)}{B(0)}}\frac{\cos \alpha_{\rm eq}}{\cos \alpha}   d\alpha_{\rm eq}\bigg ) \nonumber \\
&= 4\pi \int E^2 dE \,  \int   \sin \alpha d\alpha \, f(L,E,\sin\alpha_{\rm eq})~,
\label{eq:Jexact2}
\end{align}
which also applies to non-dipole magnetic fields.

\begin{figure}[h!]
\centering
\includegraphics[width=11 cm] {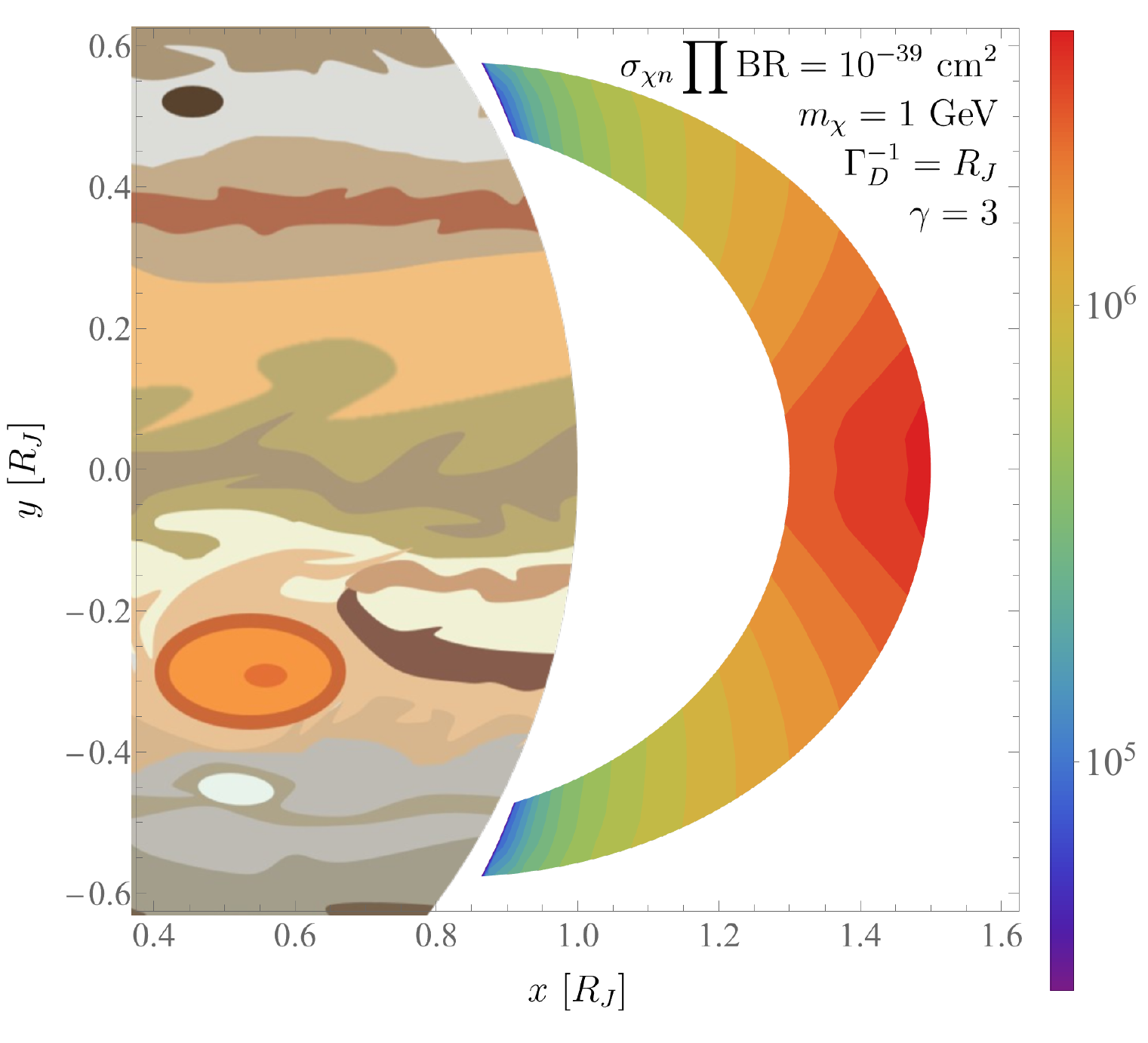}
\caption{The spatial distribution of the omnidirectional $e^\pm$ flux $J$ with an energy threshold $E_{\rm th}=10$~MeV. Presented values are based on the fully-trapped scenario and in the unit of cm$^{-2}$s$^{-1}$. We fix $m_\chi=1$~GeV, $\Gamma_D^{-1}= R_J$, $\gamma=3$, and $\sigma_{\chi n}$ $\prod $ BR = $10^{-39}$ cm$^2$.}
\label{fig:space2}
\end{figure}

In Fig.~\ref{fig:space2}, we illustrate the overall $e^\pm$ flux for a benchmark DM model in the region of $L\in[1.3,1.5]$. We assume that DM reaches an equilibrium between capture and annihilation. Then the signal rate is determined by the product of $\sigma_{\chi n}$, which determines the capture rate, and $\prod$ BR $\equiv$ BR($2\chi\to 2 \xi$) $\times$ BR($\xi\to e^+e^-$), which sets the fraction of DM annihilation resulting in the $e^+e^-$ final state. As shown in the right panel of Fig.~\ref{fig:Sync}, as $\alpha_{\rm eq}$ increases, $\tau_E|_{\rm sync}$ becomes larger, which means that the electrons could be trapped for a longer time period, resulting in a higher flux. Thus the maximum flux is observed near the equatorial region, where electrons with high-$\alpha_{\rm eq}$ tend to stay and contribute the most to $J$.

\section{Scenarios Beyond the Dipole Approximation}
\label{sec:loss}

In the previous section, the framework of electron flux calculation and major friction terms are discussed, assuming a dipole magnetic field. The characteristic time scale associated with the electron loss term is assumed to be much longer than $\tau_E|_{\rm sync}$ or $\tau_y|_{\rm sync}$. 
However, for measurements close to the Jovian atmosphere, the hard scattering/absorption by gas, plasma, or small grains could introduce a non-negligible electron loss. In addition, when $r\lesssim 1.3 R_J$, the higher multipole components of the magnetic field become significant, leading to a breakdown of the dipole approximation and electron loss~\cite{2002P&SS...50..277W}. If any of these effects take place, the loss term in Eq.~\eqref{eq:simple} could not be ignored and is assumed to take the simplified form:
\begin{equation}
\text{loss term}\simeq \tau_{\rm loss}^{-1} f~,
\end{equation}
where $\tau_{\rm loss}$ is the time scale of electron disappearances and could be a function of $L$, $E$, and $\sin\alpha_{\rm eq}$. When the electron loss is faster than the friction processes ($\tau_{\rm loss}\ll \tau_E|_{\rm sync}$), the solution of Eq.~\eqref{eq:simple} converges to $f \simeq \tau_{\rm loss} \bar{I}$ accordingly.

Depending on the level of dipole approximation violation, we consider three simplified scenarios of electron trapping in the Jovian magnetic field:
\begin{itemize}

\item {\bf Fully trapped.} The scenario corresponds to the situation in Section~\ref{sec:flux}, in which electrons are fully trapped and bounce back and forth between mirror points in the Jovian magnetosphere before they lose most of their energies, with at most minor deviations. This could apply when the observation is performed in the region of $r\gtrsim 1.3 R_J$ and near the magnetic equator~\cite{2002P&SS...50..277W}. Fully trapped electrons do not meet breaking points where the dipole approximation is badly broken when they drift along the $L$-shell. 
In addition, the region also enjoys a low gas/plasma density due to its distance from the planet. The extended halo ring~\cite{2004Icar..172...59T} affects the region, but the effect is mild~\cite{nenon2018rings}.

Provided no significant electron loss in this case, we expect that the electron lifetime is mostly determined by $\tau_E|_{\rm sync}$ as discussed in Section~\ref{sec:frictionsync}. To be conservative and take into account of various absorption terms, we still add $\tau_{\rm loss}$ in this case and approximate it as a constant in the calculation. The minimum value of $\tau_{\rm loss}$ in the fully trapped scenario is set by the largest $\tau_{\rm loss} \sim \mathcal{O}(10^{5})$~s found due to the main ring absorption~\cite{2018GeoRL..4510838N}. Strictly speaking, such a short time scale comparable to $\tau_E|_{\rm sync}$ only applies to the main ring region with $L\in [1.7,2]$ and serves as a conservative lower bound. For $\tau_{\rm loss}$ longer than $\tau_E|_{\rm sync}$, the electron absorption effect is sub-dominant. 

\item {\bf Quasi-trapped.} Electrons in this scenario could still be reflected by both mirror points along the local field lines. However, as the electron drifts along the azimuthal (longitudinal) direction due to the magnetic field's gradient, it lands on irregular field lines that lead to its removal~\cite{2002P&SS...50..277W}. More concretely, if the electron drifts from one longitude to another longitude along which the reflecting magnetic mirror points are inside Jupiter, the electrons will be absorbed by and lost to the atmosphere. In this case, $\tau_{\rm loss}$ must then be smaller than the electron drift period. The dipole approximation may still hold except for a few regions where electrons are absorbed. The azimuthal drift period $\tau_{\rm drift}$, under the dipole approximation, reads~\cite{schulz2012particle}
\begin{equation}
\label{eq:drift}
\tau_{\rm drift}\simeq 2.9\times 10^5 ~\text{s} \;  L^{-1}(0.11 \sin^2 \alpha_{\rm eq}+0.12 \sin \alpha_{\rm eq}+0.77)^{-1}\bigg(\frac{100~\text{MeV}}{E}\bigg)~. 
\end{equation}
By definition, a quasi-trapped electron or positron (with opposite drift directions) will hit at least one region it cannot pass through within its drift period. The expected time scale $\tau_{\rm loss}$ in this case is capped by $\tau_{\rm drift}/2$. More realistically, there could be multiple breaking spots for a certain $L$-shell, which further shortens $\tau_{\rm loss}$. Therefore, for this scenario, we approximate the loss time scale to be
\begin{equation}
\tau_{\rm loss} \sim \frac{\tau_{\rm drift}}{2} \times \text{const}~,
\label{eq:driftloss}
\end{equation}
where the constant $<1$ depends on the magnetic field's structure and position of the \textit{in situ} measurements. The flux predicted in this scenario is thus longitude-dependent. The precise determination of $\tau_{\rm loss}$ is highly non-trivial. Such modeling may benefit from matching with other \textit{in situ} data such as the fluxes of non-relativistic ions~\cite{2021JGRA..12628925K} or magnetic fields~\cite{2022JGRE..12707055C}. We will not pursue an accurate modeling of $\tau_{\rm loss}$ in our paper. Instead, we will only use $\tau_{\rm drift}$ multiplied by a few benchmark constants as proxies for $\tau_{\rm loss}$. Quasi-trapping could happen in regions close to the Jovian atmosphere with $r < 1.3 R_J$ and away from the magnetic equator. The electron loss time scale, $\tau_{\rm loss}$, estimated in this case could be shorter than $\tau_E|_{\rm sync}$ and sets the electron trapping time.

\item {\bf Untrapped.} In this scenario, the magnetic field lines electrons travel along are highly irregular. The field strength may even decrease (instead of increasing as in the dipole model) when the field line approaches the planet's surface. Charged particles may never be mirrored before being absorbed by the atmosphere. The electron lifetime is extremely short, and the trajectory length is characterized by the field line length outside the atmosphere, which is of $\mathcal{O}(R_J)$.

\end{itemize}

\section{Measurements and Constraints on DM Models}
\label{sec:constraints}
\begin{table}[h!]
\centering
\begin{footnotesize}
\resizebox{\textwidth}{!}{
\begin{tabular}{ccccccc}
\hline
Mission &  Instrument & Energy Range & Observable & Measured Values & Scenario\\
\hline
\multirow{2}{1.5cm}{Juno}  & ASC & $\gtrsim 10$~MeV & $J_{\rm inf}$ & $\lesssim 1\times 10^4$~cm$^{-2}$s$^{-1}$~\cite{becker2017observations}& Quasi-trapped\\
 & SRU  & $\gtrsim 10$~MeV & $J_{\rm inf}$ & $\sim 4\times 10^3$~cm$^{-2}$s$^{-1}$~\cite{becker2017observations}& Quasi-trapped\\
 \hline
\multirow{2}{1.5 cm}{Galileo Probe} & P1 & $\gtrsim 30$~MeV & $K$  & $\sim$(7 - $5\times 10^4$)~s$^{-1}$~\cite{EPIData}& Quasi-trapped \& Fully trapped$^\dagger$\\
 & P2 & $\gtrsim 100$~MeV & $K$  & $\sim$(0.2 - $1\times 10^3$)~s$^{-1}$~\cite{EPIData}&Quasi-trapped \& Fully trapped$^\dagger$\\
\hline
\end{tabular}}
\end{footnotesize}
\caption{List of \textit{in situ} measurements. The measured values of two Galileo probe channels are obtained at $L\in[1.1,1.5]$ near the magnetic equatorial region. For the two Juno RM instruments, the observations span across $L\in[1.1,1.5]$ but are away from the magnetic equator. $\dagger$ indicates that the fully trapped scenario only applies to readouts with $L\gtrsim 1.3$.}
\label{tab:instrument}
\end{table}

In this section, we will compare the relativistic electron signals with observations to derive the \textit{in situ} constraints on the DM model. There are only two missions that have probed the Jovian innermost radiation belt ($L\lesssim 1.5$): the Galileo probe~\cite{young1998galileo, 2000JGR...10512093Y} and the Juno mission~\cite{2017SSRv..213....5B}. The four \textit{in situ} measurements of electron fluxes from these two missions are summarized in Table~\ref{tab:instrument}, with their corresponding electron trapping scenarios considered. For measurements performed with $L\in[1.3, 1.5]$ and at small geomagnetic latitude $\theta_p$, the most likely scenario is that the electrons are fully trapped. For regions with $L\in[1.1, 1.3]$ and large $\theta_p$, electrons from the decays of dark mediators have smaller $\alpha_{\rm eq}$'s and a higher chance of entering the drift loss zone as described in the quasi-trapped case in the previous section. Thus we assume that for measurements performed in those regions, the quasi-trapped scenario would apply.\footnote{There is also the possibility that most electrons in those regions are untrapped. In the untrapped case, the electron lifetime would be too short $\sim 0.2$ s to give a significant flux. The best limit on $\sigma_{\chi n}\prod$BR is then about $10^{-36}$ cm$^2$ for DM mass $m_\chi$ at 1 GeV, considerably weaker than the bounds in the quasi-trapped and fully trapped scenarios, which we will discuss in detail, but still stronger than the other existing bounds for (sub)-GeV DM if the leading DM-nucleon interaction is spin-dependent. }

\begin{figure}
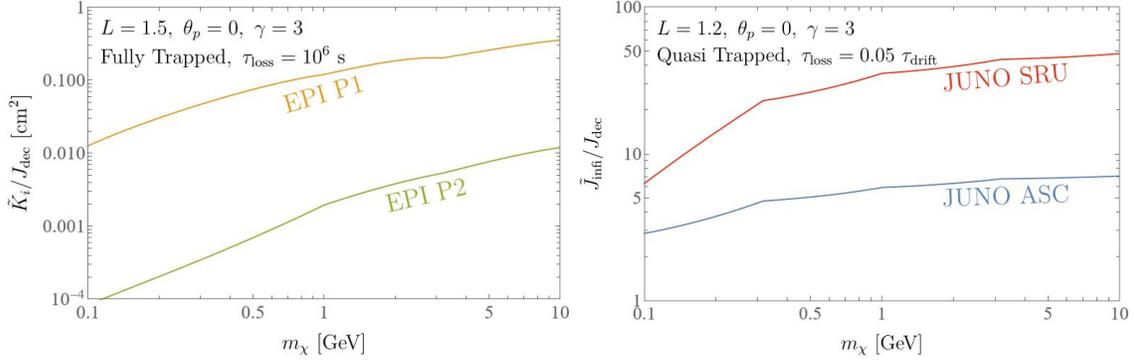

\centering
\includegraphics[height=4.8 cm]{Geometric_1}
\includegraphics[height=4.8 cm]{Geometric_2}
\caption{Relations between the predicted $e^\pm$ flux from mediator decays $J_{\rm dec}$ and corresponding observables as a function of $m_\chi$. Both plots are computed with benchmark values of $\gamma=3$ and $\theta_p=0$. \textbf{LEFT:} the ratio between $\tilde{K}_i$ and $J_{\rm dec}$ for the Galileo probe channels P1 and P2 at $L=1.5$ assuming $\tau_{\rm loss}=10^6$~s, which could be interpreted as the average effective geometric factor. \textbf{RIGHT:} the ratio between the inferred flux $\tilde{J}_{\rm inf}$ and the DM induced flux $J_{\rm dec}$, which could be understood as the ratio of the average geometric factors for $e^\pm$ in DM and background models, for the Juno RM instruments SRU and ASC at $L=1.2$, respectively. We take $\tau_{\rm loss}=0.05 \tau_{\rm drift}$.}
\label{fig:Accept}
\end{figure}

Detectors carried by the Galileo probe~\cite{fischer1996high} and Juno orbiter~\cite{becker2017observations} do not provide spectra for sub-GeV-scale $e^\pm$. Instead, the readouts are related to their geometric factors, $F(E)$'s, describing the effective areas that detectors can receive the electron signals. Harder electrons will have larger $F(E)$'s in general due to their high shield-penetrating efficiencies.\footnote{Since these detectors do not utilize magnetic fields to distinguish different charges, we take $F(E)$'s for electrons and positrons to be identical.} The $F_i(E)$ ($i=$ P1, P2) for the Galileo EPI channels are reported in~\cite{nenon2018physical}. The observed count rates $K_i$'s (number of electrons recorded per second) could be found in~\cite{EPIData} for P1 and P2 channels with different energy ranges, which are listed in Table~\ref{tab:instrument}. In our DM model, the predicted count rates $\tilde{K}_i$'s are given by
\begin{align}
\tilde{K}_i(L,\theta_p) &=  4\pi \int  \sin\alpha d\alpha  \int E^2 d E \, F_i(E)  f(L,E,\sin\alpha_{\rm eq}) =  \int   F_i(E) \left(\frac{\partial J}{\partial E}\right)_{\rm dec} dE ~, \nonumber \\
&= J_{\rm dec} \int   F_i(E) J_{\rm dec}^{-1} \left(\frac{\partial J}{\partial E}\right)_{\rm dec} dE~,
\end{align}
where we use Eq.~\eqref{eq:Jexact2} in the second equality; $\left(\partial J/\partial E\right)_{\rm dec}$ is the normalized differential energy spectrum of the electrons from dark mediator decays, and the total flux $J_{\rm dec} = \int dE \left(\partial J/\partial E\right)_{\rm dec}$. Requiring $\tilde{K}_i \leq K_i$, we could set constraints on DM models, in particular, the combination $\sigma_{\chi n} \prod {\rm Br}$. Note that this approach is conservative by being agnostic about the contributions to $K_i$ from astrophysical Jovian electrons and other charged particles, which constitute the backgrounds for our DM search. Currently, such backgrounds are not well understood. With further developments in understanding the Jovian radiation belts, we could hope for stronger constraints on the DM parameters and even hints for new physics if anomalous features were observed, which could not be explained by astrophysical sources. Choosing a DM benchmark model with $\gamma=3$, we plot the ratio between the count rates induced by unit total flux, $\tilde{K}_i/J_{\rm dec}$, at the magnetic equator with a benchmark $\tau_{\rm loss}=10^6$~s in the left panel of Fig.~\ref{fig:Accept}. When model parameters and positions of measurements vary, the spectrum $(\partial J/\partial E)_{\rm dec}$ also changes and gives rise to similar but different values of $\tilde{K}_i/J_{\rm dec}$. This ratio indicates the averaged effective geometric factor weighted by the spectrum in the DM model. The figure shows that $\tilde{K}_i/J_{\rm dec}$ increases with $m_\chi$ or equivalently with the electron energy, as expected.

For Juno's RM data collected by the Advanced Stellar Compass (ASC) and SRU, only the inferred omnidirectional fluxes with $E>10$~MeV, which we denote as $J_{\rm inf}$, are available and listed in Table~\ref{tab:instrument}. Such inferred fluxes assume an input spectrum shape of astrophysical electrons around Jupiter, denoted as $(\partial J/\partial E)_{\rm bkg}$, which is not necessarily normalized~\cite{becker2017juno}. The spectrum of the Jovian electrons is much softer than the DM-induced one. The observed count rates $K_i$, $i=$ASC, SRU, which are not published, are related to $J_{\rm inf}$ as 
\begin{equation}
K_i=   \frac{J_{\rm inf}}{J_{\rm bkg}} \int_{10~\text{MeV}}^\infty F_i(E) \left(\frac{\partial J}{\partial E}\right)_{\rm bkg} d E~,
\end{equation}
where $J_{\rm bkg} = \int \left(\frac{\partial J}{\partial E}\right)_{\rm bkg} d E$ and $F_i(E)$ encodes the detector efficiency. The ratio $J_{\rm inf}/J_{\rm bkg}$ fixes the normalization of the spectrum. 
By matching the count rates for each probe, the flux $J_{\rm dec}$ induced by the dark mediator decays could be translated to a predicted inferred flux $\tilde{J}_{\rm inf}$ as:
\begin{align}
& \int_{10~\text{MeV}}^\infty F_i(E) \left(\frac{\partial J}{\partial E}\right)_{\rm dec} d E = \frac{\tilde{J}_{\rm inf}}{J_{\rm bkg}} \int_{10~\text{MeV}}^\infty F_i(E) \left(\frac{\partial J}{\partial E}\right)_{\rm bkg} d E 
 \nonumber \\
& \Rightarrow \quad \tilde{J}_{\rm inf} = J_{\rm bkg} \frac{\int_{10~\text{MeV}}^\infty F_i(E) \left(\frac{\partial J}{\partial E}\right)_{\rm dec} d E} {\int_{10~\text{MeV}}^\infty F_i(E) \left(\frac{\partial J}{\partial E}\right)_{\rm bkg} d E} = J_{\rm dec}\frac{\int_{10~\text{MeV}}^\infty F_i(E) J_{\rm dec}^{-1}\left(\frac{\partial J}{\partial E}\right)_{\rm dec} d E} {\int_{10~\text{MeV}}^\infty F_i(E) J_{\rm bkg}^{-1}\left(\frac{\partial J}{\partial E}\right)_{\rm bkg} d E} ~.
\end{align} 
We could then set constraints on DM parameters by requiring $\tilde{J}_{\rm inf} \leq J_{\rm inf}$. 
The right panel of Fig.~\ref{fig:Accept} plots the ratio between $\tilde{J}_{\rm inf}$ and $J_{\rm dec}$ for different $m_\chi$'s. Although the injected $e^\pm$ spectrum gets softened by the synchrotron radiation, the average geometric factor of electrons from dark mediator decays exceeds that of the background, leading to $\tilde{J}_{\rm inf}/J_{\rm dec}>1$ and higher sensitivities for any $m_\chi> 100$~MeV.

\begin{figure}[h!]
\centering
\includegraphics[width=7.5 cm] {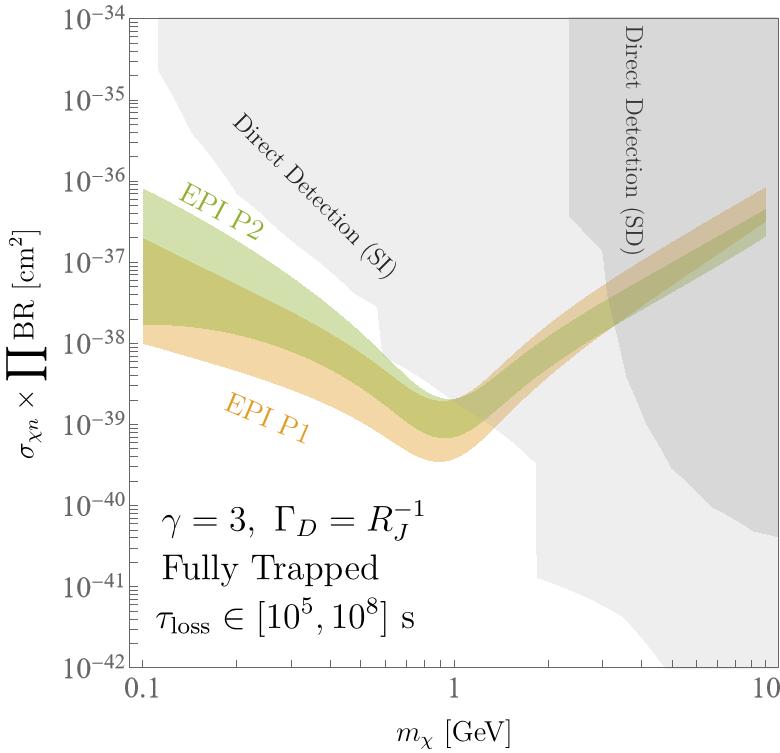}
\includegraphics[width=7.5 cm] {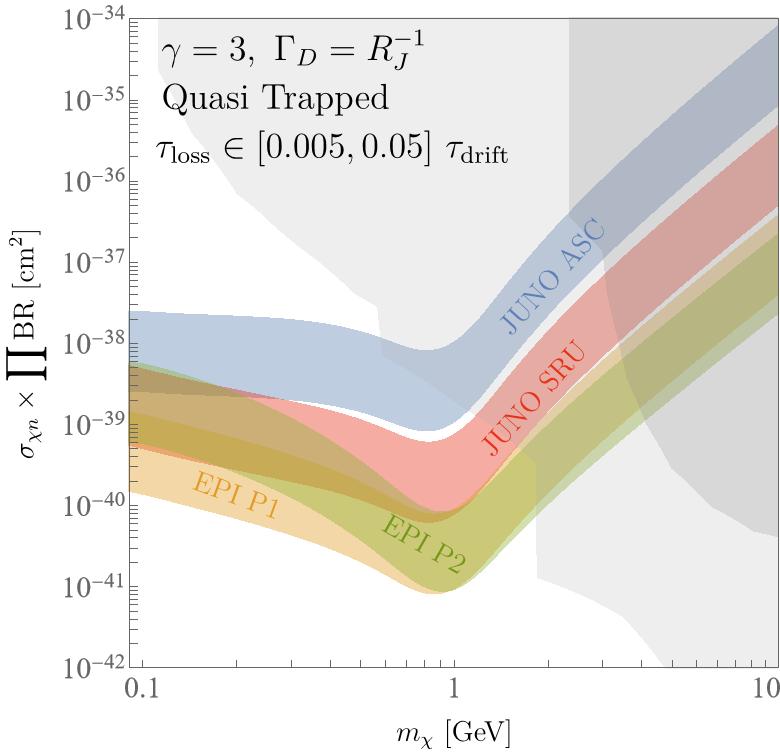}
\caption{Observed upper bounds on $\sigma_{\chi n} \times \prod$ Br due to the relativistic $e^\pm$ flux from dark mediator decays in different trapping scenarios. \textbf{LEFT:} limits from the Galileo probe EPI P1 and P2 observations in the fully trapped scenario ($L\simeq 1.45$). The band for each channel is obtained by varying $\tau_{\rm loss} \in[10^5,10^8]$~s (corresponding to upper and lower ends of the band). \textbf{RIGHT:} limits from the two Galileo EPI channels as well as Juno ASC and SRU measurements in the quasi-trapped scenario at $L\lesssim 1.2$. Since the exact time scale of $\tau_{\rm loss}$ in this case is unknown, we show the band from computing $\tau_{\rm loss}$ using Eq.~\eqref{eq:drift} and~\eqref{eq:driftloss} with the constant $\in [0.01,0.1]$ (leading to upper and lower band edges). The lighter (darker) grey regions are constraints on $\sigma_{\chi n}$ from direct detection experiments, assuming SI (SD) scattering~\cite{XENON:2018voc,CDEX:2019hzn,XENON:2019zpr, PandaX-4T:2021bab, LUX-ZEPLIN:2022qhg, Behnke:2016lsk,SuperCDMS:2017nns, PICO:2019vsc}. }
\label{fig:Projectionhits}
\end{figure}

As discussed in Section~\ref{sec:capture} and~\ref{sec:model}, when the equilibrium between DM capture and annihilation is reached, the $e^\pm$ flux in the DM model is entirely determined by $\sigma_{\chi n} \prod$ BR. We first consider the constraints on this combination in the fully trapped scenario, which applies to the Galileo EPI probes P1 and P2 with readouts at $L\gtrsim 1.3$ near the magnetic equator. Constraints are calculated by comparing $\tilde{K}_i$'s induced by $\xi$ decays with the observed count rates. The values of $L$ and $\theta_p$ for each observation are obtained using the dipole component of the JRM09 magnetic model~\cite{connerney2018new} for simplicity and concreteness. As discussed in Section~\ref{sec:loss}, the constraints in the fully trapped case are not very sensitive to $\tau_{\rm loss}$ since electron loss in this region is a sub-dominant effect. The results are presented in the left panel of Fig.~\ref{fig:Projectionhits}, with the bands obtained by varying $\tau_{\rm loss}$ from $10^{5}$ to $10^{8}$~s. We fix $\gamma=3$, and $\Gamma_D=R_J^{-1}$ so that the proper lifetime of dark mediator $\xi$ is 0.2 s. As $m_{\chi}$ increases, $\tau_E|_{\rm sync}$ becomes more dominating, reducing the differences in bounds introduced by a varying $\tau_{\rm loss}$ and making the band narrower. For a given $\tau_{\rm loss}$, the bounds are the strongest, $\sigma_{\chi n} \prod$ Br $\lesssim {\cal O}(10^{-39})$ cm$^2$, when $m_\chi \simeq 1$~GeV, where the DM capture is efficient and injected electrons are energetic. Direct detection bounds on both spin-independent (SI)~\cite{XENON:2018voc,CDEX:2019hzn,XENON:2019zpr, PandaX-4T:2021bab, LUX-ZEPLIN:2022qhg} and spin-dependent (SD) DM-nucleon scattering cross sections~\cite{Behnke:2016lsk,SuperCDMS:2017nns,XENON:2019zpr,PICO:2019vsc} are also presented for comparison.\footnote{Cosmological bounds on $\sigma_{\chi n}$ in this mass range, which are not plotted, are weaker~\cite{Gluscevic:2017ywp, Buen-Abad:2021mvc}.} We see that the Jupiter \textit{in situ} electron flux measurements set stronger constraints for $m_\chi$ below a few GeV, in particular, if the interaction between DM and nucleons is spin-dependent. Note that SD and SI scatterings do not make a difference for DM capture inside Jupiter since Jupiter is mostly made up of hydrogen atoms.

The next step is to consider measurements in regions with $L\lesssim 1.3$ where the electrons are quasi-trapped. The method is similar to the fully-trapped case discussed above but with $\tau_{\rm loss}$ determined by drift period as in Eq.~\eqref{eq:drift} and~\eqref{eq:driftloss}. In particular, we vary the constant in Eq.~\eqref{eq:driftloss} between $0.01$ to $0.1$ as benchmark values. The chosen range of the constant is arbitrary. It should depend on the measurement position including the longitude and becomes larger for smaller $L$. Nevertheless, the simplification allows us to estimate the DM parameter space Jupiter measurements could probe. We will leave a more precise determination to future work. The constraints from the Galileo probe data are shown in the right panel of Fig.~\ref{fig:Projectionhits}. Again we fix $\gamma=3$ and $\Gamma_D=R_J^{-1}$. In the quasi-trapped case, $\tau_{\rm loss}$ is shorter than that in the fully-trapped case, and the electron loss becomes important for evaluating the electron trapping time. However, the Galileo EPI readouts are much smaller in the innermost region, which compensates for the shorter $\tau_{\rm loss}$, resulting in stronger limits, $\sigma_{\chi n} \times \prod$ Br $\lesssim {\cal O} (10^{-41})$ cm$^2$ at $m_\chi$ around 1 GeV.  Limits from the two Juno RM investigations are also included in the right panel of Fig.~\ref{fig:Projectionhits}, obtained by comparing $\tilde{J}_{\rm inf}$ with $J_{\rm inf}$ reported in~\cite{becker2017observations}. Their sensitivities are weaker compared to the Galileo ones, limited by multiple factors such as starlight backgrounds and their dynamic ranges~\cite{becker2017juno}.

\begin{figure}[h!]
\centering
\includegraphics[width=7.5cm]{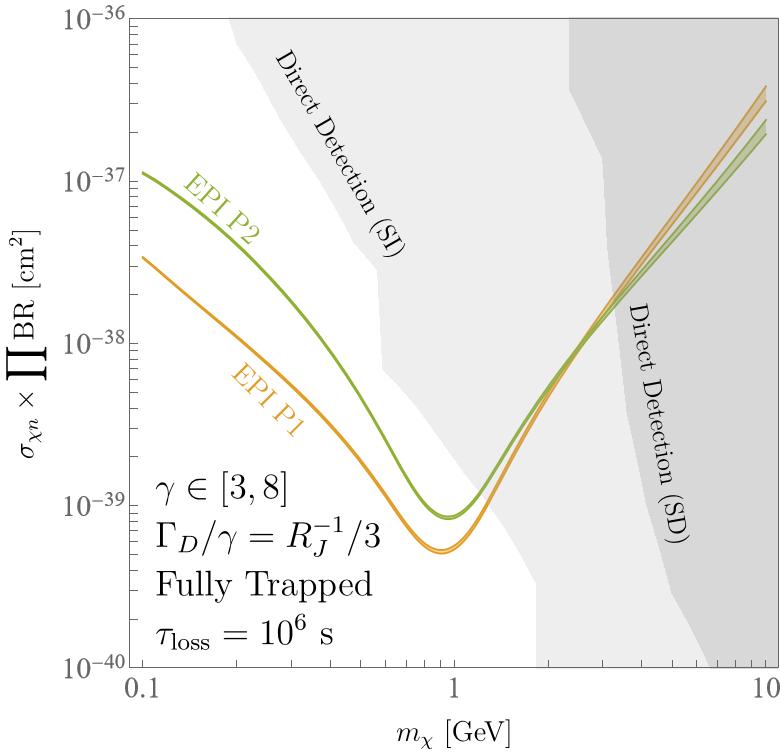}
\includegraphics[width=7.5cm]{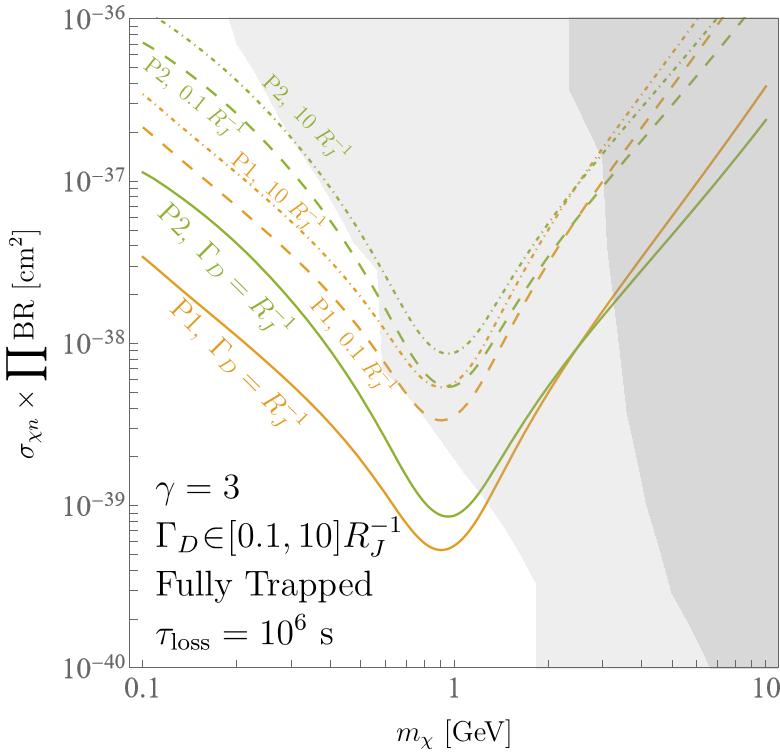}
\caption{Similar to the left panel of Fig.~\ref{fig:Projectionhits} but with different model parameters. The electron loss time scale $\tau_{\rm loss}=10^{6}$~s is applied in both panels. \textbf{LEFT: } Comparison between limits with $\gamma=3$ (upper band edge) and $\gamma=8$ (lower band edge), while the ratio $\Gamma_D/\gamma$ is fixed at $R_J^{-1}/3$. \textbf{RIGHT: } Comparison between limits from different $\Gamma_D$'s. Solid curves are results from $\Gamma_D=R_J^{-1}$, while dashed (dot-dashed) curves are results with $\Gamma_D= 0.1 R_J^{-1} (10 R_J^{-1})$. }
\label{fig:Projectionhits2}
\end{figure}

Since the constraints in Fig.~\ref{fig:Projectionhits} are obtained with the benchmark $\gamma=3$ and $\Gamma_D=R_J^{-1}$, it is necessary to check limits with alternative model parameters. Here we use the fully-rapped scenario constraints with $\tau_{\rm loss}=10^{6}$~s as the representative. In the left panel of Fig.~\ref{fig:Projectionhits2}, we show the comparison between models with $\gamma=3$ and $\gamma=8$. To keep the proper decay length of $\xi$ the same for a fair comparison, we fix $\Gamma_D/\gamma =R_{J}^{-1}/3$ in both cases. Consequently, the difference between the two benchmarks is only induced by the injection term $\bar{I}$ and is minor, as shown in the figure. The effect of varying $\Gamma_D$ could be estimated with Eq.~\eqref{eq:decaydensity} and is presented in the right panel of Fig~\ref{fig:Projectionhits2}. For $\Gamma_D/\gamma \lesssim R_J^{-1}$, the smaller $\Gamma_D$ is, the fewer decays happen, resulting in weaker constraints. When $\Gamma_D/\gamma \gg R_J^{-1}$, most decays happen inside Jupiter, and the bounds get weaker as well.

Finally, we comment on the assumption of the equilibrium between DM capture and annihilation. We focus on the DM mass range ${\cal O} (0.1 - 10)$ GeV and show that Jupiter missions could constrain $\sigma_{\chi n} \prod$ Br of ${\cal O}(10^{-41}-10^{-39})$ cm$^2$ under different assumptions. From Eq.~\eqref{eq:teq}, we see that the equilibrium time scale is shorter than or comparable to the Jupiter age for DM about or heavier than 1~GeV. Thus the equilibrium assumption shall be valid. In the region of $m_\chi \lesssim 1$~GeV, the evaporation effect could become significant, and the limits we plot are only tentative. We leave a full analysis including both annihilation and evaporation in this regime to future work.

\section{Summary and Outlook}
\label{sec:con}

In this article, we propose an intriguing connection between studies of Jovian radiation belts and DM searches. We consider a class of DM models in which DM captured by Jupiter could annihilate into a pair of long-lived dark mediators, which subsequently decay into $e^\pm$ outside Jupiter. This could happen when the decay length of the mediator is comparable to the Jupiter radius, e.g., a sub-GeV scale dark photon with kinetic mixing parameter $\epsilon \sim 10^{-10}$ which is not (fully) covered experimentally. The produced $e^\pm$ could be either fully- or quasi-trapped in the innermost radiation belts and contribute to energetic electron fluxes, which are recorded in the \textit{in situ} measurements by various Jupiter missions. We apply the data collected by the Galileo probe and Juno missions and find powerful constraints on the product of $\sigma_{\chi n}$, DM-nucleon scattering cross section, and $\prod$ Br, the branching fraction of DM annihilations ending in $e^\pm$ final state, for DM mass between (0.1 - 10) GeV. In particular, for DM at 1 GeV, the bound on the product could be as strong as $10^{-41}$~cm$^2$ from the data collected by Galileo EPI probes at $L \lesssim 1.2$ where the electrons are quasi-trapped. The quasi-trapping case is subject to uncertainties related to the electron loss effects, which we adopt a simplified modeling. A weaker but potentially more reliable constraint could be derived with data collected at larger $L$ ($L \approx 1.5$) where the electrons are fully-trapped and the electron loss is sub-dominant. In this case, the upper limit on $\sigma_{\chi n} \times \prod$ Br is around $10^{-39}$ cm$^2$ for DM at 1 GeV. These bounds could be comparable to or stronger than current GeV-scale DM direct detection searches, in particular, for spin-dependent DM-nucleon scattering.

Our study is only an initial effort to apply the Jovian data from the past, ongoing, and future Jupiter missions to probe new physics beyond the standard model of particle physics. The analysis could be improved in several aspects: the investigation in the quasi-trapping scenario could be refined with a more precise modeling of the electron loss effects; a better understanding of the astrophysical electron sources could allow us to set stronger constraints on the DM parameters. We also want to stress that future Jupiter missions may enable more precise measurements of energetic electron fluxes and the corresponding spectra at different positions~\cite{2021ExA...tmp..136R}, which could greatly strengthen the bound. Beyond the analysis we did, we list and outline several other (but not all) exciting avenues to explore the power of Jovian data in new physics searches below.

\begin{itemize}
\item {\bf X-rays from inverse Compton scattering} The electrons from dark mediator decays could produce X-rays through inverse Compton scattering with solar photons.\footnote{We thank Elias Roussos for pointing out this possibility.}  The interaction rate between injected electrons and solar photons is low, leaving the energy loss due to this effect negligible in calculations of electron's flux. However, photons back-scattered by ultra-relativistic MeV-GeV-scale electrons pick up energies in the keV-MeV range, depending on the electron spectrum. So far, there are no \textit{in situ} measurements of Jovian $X$-rays. Nevertheless, there are data sets on the Jovian X-ray backgrounds from Chandra, XMM-Newton and Suzaku X-ray telescopes~\cite{bhardwaj2005solar,elsner2005simultaneous,branduardi2007latest,branduardi2008spectral,dunn2020jupiter}, which are all located near the Earth. In particular, in~\cite{Ezoe:2010hw}, it was pointed out that a diffuse hard X-ray (1-5~keV) emission around Jupiter could be explained by solar photons scattering with a large population of energetic electrons trapped within $r\simeq (4$-$8) R_J$. A quick estimate shows that such a large electron source might be provided by injected electrons from DM annihilations in our model. We will leave a more detailed study for the future.

There are other secondary conversion products that could be observational targets, such as ions from electron impact ionization or even higher energy photons, i.e., $\gamma$ rays from electromagnetic radiation~\cite{Leane:2022TBA} or electron-positron annihilations.

\item {\bf Positron signals} The \textit{in situ} measurements discussed in this paper are implemented by instruments without magnetic fields and do not distinguish positrons from electrons. Since Jupiter is not a known active positron source, a Jovian positron signal could be striking if detected. One possibility is that the high-energy positrons outside Jupiter predicted in our DM model could escape the magnetosphere and transfer to the inner heliosphere through the twisted magnetic field lines, namely the Parker spiral~\cite{parker1958dynamics}. Then one could use PAMELA~\cite{PAMELA:2011bbe} or AMS~\cite{AMS:2014xys} cosmic-ray detectors to study the Jovian positrons near the earth. The relative motion between Jupiter and the Earth will also create a $\sim 13$ month period of positron flux. Such dynamics and detection potential have been studied for the Jovian electrons transported to the earth orbit~\cite{potgieter2002effects, ferreira2005transport, DIFELICE20082037, strauss2011modeling,vogt2018jovian,vogt2022numerical}.

\item {\bf Solar axion conversion to X-rays} This possibility is not related to the DM model in this paper.\footnote{We thank Ben Monreal for bringing up this possibility.} It is well known that the Sun could produce axions, one of the most motivated feebly-coupled particles beyond the standard model, with energies set by the solar core temperature, which is in the keV range~\cite{Fukugita:1982ep, Sikivie:1983ip}. These relativistic axions from the Sun, also referred to as solar axions, have become a standard benchmark scenario for direct detection experiments~\cite{Arisaka:2012pb, XENON100:2014csq, LUX:2017glr, PandaX:2017ock, XENON:2020rca}. Jupiter, with its strong magnetic field and huge volume, could serve as a giant cavity for oscillations between solar axions and photons through the axion-photon coupling. The signal we could look for is that the solar axions propagate through Jupiter, and some of them convert into X-rays in the keV range at the dark side of Jupiter away from the Sun. A similar idea was proposed a while back to use the Earth as the converter~\cite{Davoudiasl:2005nh,Davoudiasl:2008fy}. The estimated conversion probability (scaling as $B^2 R^2$ with $B$ the magnetic field strength and $R$ the distance axion travels) for the Earth could be comparable to the CERN axion solar telescope (CAST), a leading terrestrial experiment setting strong bound on axion-photon coupling~\cite{CAST:2004gzq, CAST:2007jps, CAST:2008ixs, CAST:2011rjr, CAST:2013bqn}. While the solar axion flux at Jupiter is reduced by a factor of 25 compared to that at the Earth since Jupiter is further away from the Sun, the conversion probability is larger since the Jovian $B$ is one order of magnitude above the Earth one and the size of the magnetic field the axion could traverse is also larger. The challenges are two-fold: {\it i)} there is no current \textit{in situ} measurement of X-ray at the dark side of Jupiter, as mentioned in the first item; {\it ii)} the Jovian X-ray background close to the planet and in regions at low latitudes away from the pole aurorae is not fully understood. 

\end{itemize}

\section*{Acknowledgements}

We thank Heidi Becker, Raghuveer Garani, Peter Kollmann, Rebecca Leane, Matt Reece, Elias Roussos, and Hai-Bo Yu for useful communication and discussions. JF and LL are supported by the DOE grant DE-SC-0010010, the NASA grant 80NSSC18K1010 and 80NSSC22K081. 

\appendix

\section{Monte Carlo Simulations of the Injected Electron Phase Space Distribution}
\label{app:simulations}

The injection distribution $I$ depends on $E$ and $\sin \alpha$ in a highly nontrivial way, due to the complicated $e^\pm$ energy spectrum from boosted mediator decays. In contrast to the isotropic electron injection from slow neutron decays~\cite{1961JGR....66.4027L}, the energy spectrum of injected electrons is strongly correlated with the distribution of the pitch angle $\alpha$. The analytical form of $I$ also contains divergences near the boundary of kinematically allowed regions. In order to get the integrated form of $I$, 
we adopt the Monte Carlo approach that is relatively stable numerically and computationally cheap.

To start with, we generate an $e^\pm$ sample sets $\{D\}={D_1(w_1),D_2(w_2),...,D_n(w_n)}$ of the same $L$-shell. Each point $D_i$ is obtained by sampling the boosted two-body phase space of a mediator decaying at a distance $r_i$ and geomagnetic latitude $\theta_{p,i}$. The pitch angle $\alpha_i$ and $\alpha_{{\rm eq},i}$ for each $D_i$ are calculated using the magnetic dipole field model, while its energy $E_i$ is directly known from the decay kinematics. For each $D_i$, a weight $w_i$ is assigned. The Monte Carlo sample $\{D\}$ would be a representative of the injection term when it satisfies:
\begin{equation}
 \sum_{\mathclap{\substack {S_i \in [S,S+dS], \\ E_i \in[E,E+d E], \\ \alpha_i \in[\alpha,\alpha+d\alpha]}}} w_i \simeq  4\pi I(L,E,\sin\alpha)E^2  dE \, \sin\alpha d\alpha \, dS~.
\end{equation}
Integrating the equation above over $\alpha$ and $E$ and plugging it into Eq.~\eqref{eq:rhoD}, we have:
\begin{equation}
\sum_{\mathclap{\theta_{p,i} \in [\theta_p,\theta_p+d\theta_p]}} w_i= \rho_D(r) \frac{dS}{d\theta_p} d\theta_p \simeq \frac{\rho_D(L)}{\cos^4 \theta_p} \frac{dS}{d\theta_p} d\theta_p~,
\end{equation}
where we take the conservative approximation of $\rho_D(r)> \frac{\rho_D(L)L^2}{r^2}\frac{\rho_D(L)}{\cos^4 \theta_p} $ to make the calculation $\Gamma_D$ invariant. Since our $\{D\}$ is obtained by evenly sampling $\theta_p\in[-\arccos\sqrt{L^{-1}},\arccos\sqrt{L^{-1}}]$, we can thus assign
\begin{equation}
w_i = \frac{2\arccos\sqrt{L^{-1}}}{N_{\rm sample}} \frac{\rho_D(L)}{\cos^4 \theta_p}\frac{dS}{d\theta_p}~,
\end{equation}
where $N_{\rm sample}$ is the total number of sample points. 

To calculate the $L$-shall average $\bar{I}$, we follow the standard steps in~\cite{1961JGR....66.4027L}:
\begin{align}
\int_z I dz  E^2 dE \sin\alpha_{\rm eq} d\alpha_{\rm eq}  &= \int_S I \frac{dS}{\cos\alpha}  E^2 dE \sin\alpha_{\rm eq} d\alpha_{\rm eq} = \int_S I \frac{dS}{\cos\alpha} \frac{d \cos \alpha_{\rm eq}}{d \cos \alpha}  E^2 dE \sin\alpha d\alpha \nonumber \\
& = \int_S I dS \frac{B(0)}{B(\theta_p)}\frac{1}{\cos\alpha_{\rm eq}}  E^2 dE \sin\alpha d\alpha \nonumber  \\
&\simeq \frac{1}{4\pi}\sum_{\substack{E_i \in[E,E+d E], \\ \alpha_{i} \in[\alpha,\alpha+d\alpha]}} \frac{B(0)}{B(\theta_p) \cos \alpha_{\rm eq}} w_i~.
\end{align}
Finally, the trajectory length 
\begin{equation}
\int_z dz = \int_{-\arcsin\sqrt{L^{-1}}}^{\arcsin\sqrt{L^{-1}}} \frac{dS}{d\theta_p} \frac{1}{\cos \alpha} d\theta_p \equiv 2 T(\sin\alpha_{\rm eq})~.
\end{equation}
Although $T$ has no analytical form in general, there is a well-known numerical approximation~\cite{schulz2012particle}
\begin{equation}
T(y)\simeq 1.3802 -0.3198 (y+\sqrt{y})~.
\end{equation}

\bibliography{Ref}
\end{document}